\documentclass[unnumsec,webpdf,modern,large]{oup-authoring-template}%
\usepackage{natbib}

\theoremstyle{thmstyleone}%
\theoremstyle{thmstyletwo}%
\theoremstyle{thmstylethree}%

\usepackage{tikz}
\usepackage{pgfplots}
\pgfplotsset{compat=1.18} 
\usepackage{soul}
\usepackage[nolist]{acronym}
\usepackage{booktabs, multirow, array}
\usepackage{longtable}
\usepackage{float}
\makeatletter
\renewcommand*\l@subsubsection[2]{{}{}{}} %
\makeatother
\usepackage{subcaption}%
\usepackage{booktabs}
\usepackage{tabularx}

\newcounter{mypromptctr}

\usepackage{textcomp, siunitx}
\AtBeginDocument{
  \catcode`_=12 
}
\usepackage{xcolor}
\usepgfplotslibrary{groupplots}
\definecolor{oiBlue}{RGB}{0,114,178}

\definecolor{llgray}{gray}{0.95}
\newcolumntype{P}[1]{>{\raggedright\arraybackslash}p{#1}}

\usepackage[most]{tcolorbox}

\usepackage{cleveref}

\newcommand{\mail}[1]{\href{mailto:#1}{#1}}

\begin{document}

\journaltitle{Under review}
\DOI{DOI HERE}
\copyrightyear{2025}
\pubyear{2025}
\access{Advance Access Publication Date: Day Month Year}
\appnotes{Paper}

\title[Improving Phishing Resilience with AI-Generated Training]{Improving Phishing Resilience with AI-Generated Training: Evidence on Prompting, Personalization, and Duration}

\author[1]{Francesco Greco\ORCID{0000-0003-2730-7697}}
\author[1,$\ast$]{Giuseppe Desolda\ORCID{0000-0001-9894-2116}}
\author[1]{Cesare Tucci\ORCID{0000-0001-5181-7115}}
\author[1]{Andrea Esposito\ORCID{0000-0002-9536-3087}}
\author[1]{Antonio Curci\ORCID{0000-0001-6863-872X}}
\author[1]{Antonio Piccinno\ORCID{0000-0003-1561-7073}}

\authormark{Greco F. et al.}

\corresp[$\ast$]{Corresponding author. \mail{giuseppe.desolda@uniba.it}}
\address[1]{%
      \orgdiv{Department of Computer Science},
      \orgname{University of Bari Aldo Moro},
      \orgaddress{%
      \street{Via E. Orabona 4},
      \postcode{70125},
      \country{Italy}
      }
}

\received{Date}{0}{Year}
\revised{Date}{0}{Year}
\accepted{Date}{0}{Year}

\abstract{%
Phishing remains a persistent cybersecurity threat; however, developing scalable and effective user training is labor-intensive and challenging to maintain. Generative Artificial Intelligence offers an interesting opportunity, but empirical evidence on its instructional efficacy remains scarce.
This paper provides an experimental validation of Large Language Models (LLMs) as autonomous engines for generating phishing resilience training. Across two controlled studies ($N=480$), we demonstrate that AI-generated content yields significant pre-post learning gains regardless of the specific prompting strategy employed. Study~1 ($N=80$) compares four prompting techniques, finding that even a straightforward \textit{``direct-profile'' strategy---simply embedding user traits into the prompt---}produces effective training material. Study~2 ($N=400$) investigates the scalability of this approach by testing \emph{personalization} and \emph{training duration}. Results show that complex psychometric personalization offers no measurable advantage over well-designed generic content, while longer training duration provides a modest boost in accuracy. These findings suggest that organizations can leverage LLMs to generate high-quality, effective training at scale without the need for complex user profiling, relying instead on the inherent capabilities of the model.}


\maketitle

\begin{acronym}
    \acro{AI}{Artificial Intelligence}
    \acro{HCAI}{Human-Centered Artificial Intelligence}
    \acro{HCD}{Human-Centered Design}
    \acro{HCI}{Human Computer Interaction}
    \acro{UI}{User Interface}
    \acro{EU}{European Union}
    \acro{LLM}{Large Language Model}
    \acro{RQ}{Research Question}
    \acro{PETA}{Phishing Education, Training, and Awareness}
    \acro{EI}{Emotional Intelligence}
\end{acronym}

\section{Introduction}

Phishing is one of the most pervasive and damaging cyber threats worldwide. Industry reports indicate that social engineering attacks remain the primary entry point for threats, with phishing (including vishing, SMishing, and malvertising) accounting for up to 80\% of cases ~\citep{apwg2024trends, ibm2024breach, ENISA2025}. This high percentage of phishing effectiveness demonstrates that automatic filtering techniques (e.g., blocklists, machine learning-based detection) \citep{Basit2021Survey} and warning dialogs shown to users in cases of suspicious emails \cite{buono2023warnings, Desolda2025Apollo, desolda2023explanations} are insufficient to defend users from this threat. 

Several factors contribute to the effectiveness of phishing attacks. Some studies demonstrated that many users focus on the visual appearance and content of emails and web pages, while ignoring more reliable security indicators~\citep{dhamija2006phishing, alsharnouby2015phishing, Gallo2024HumanFactors}. This implies that addressing merely technical defense techniques is insufficient to mitigate phishing effectively, since it requires addressing human factors through usable security mechanisms and training~\citep{Desolda2022PhishingSLR,khonji2013phishing}. 
In response, researchers and practitioners have proposed a wide range of \ac{PETA} programs, including embedded training and game-based modules, as well as simulated phishing campaigns and organizational awareness programs~\citep{ sheng2010falls, canfield2016training, reinheimer2020investigation, zielinska2014phishing}.

While these approaches can reduce susceptibility, they also suffer from significant limitations. First, \emph{creating and maintaining high-quality training content is labor-intensive} because designers must craft realistic e-mails, scenarios, explanations, and feedback messages that reflect current attack techniques and organizational practices~\citep{reinheimer2020investigation}. 
Keeping these materials up to date as attackers rapidly evolve their tactics imposes a continuous and high effort on security teams, which is difficult to sustain at scale~\citep{khonji2013phishing,reinheimer2020investigation}. 
Second, many programs adopt a \emph{"one-size-fits-all" approach}, delivering the same content to heterogeneous user populations with different roles, expertise, prior knowledge, and psychological profiles~\citep{sheng2010falls,canfield2016training}. This might limit their ability to adapt to individuals who may require different examples, explanations, or motivational strategies.
Third, evidence suggests that the impact of PETA interventions can be \textit{uneven and short-lived}. Several studies report initial reductions in click rates or improved quiz performance, followed by decay over time or mixed results across user groups ~\citep{sheng2010falls,canfield2016training,reinheimer2020investigation}. 
Timing and frequency also matter: training that is too infrequent, too generic, or poorly aligned with users' daily tasks may fail to translate into durable changes in behavior~\cite{reinheimer2020investigation}. 
Finally, most existing \ac{PETA} programs provide limited support for \textit{rapid experimentation} on training design choices (e.g., how content structure, personalization, or duration shape learning), because each new variant requires substantial manual authoring and coordination.

\acp{LLM} offer a promising way to rethink how phishing-awareness training is produced and delivered. They can help generate fluent, contextually rich text and dynamic content in various styles or tailored for different personas based on natural language prompts~\citep{brown2020, Heiding2023}. 
They may also facilitate the automatic creation of realistic phishing examples, explanatory materials, and interactive training modules tailored to specific topics or user profiles, significantly reducing the manual effort required compared to traditional methods ~\citep{Bethany2025LateralPhishing, Qi2025SpearBot}.
However, this opportunity raises new questions: \emph{Which prompting strategies yield effective training content?}, \emph{Does LLM-based personalization meaningfully improve learning?}, and \emph{How much training is enough, and does longer always mean better?}

In this paper, we take a first step towards answering these questions by conducting two controlled experiments that systematically evaluate LLM-generated phishing-awareness training. 
Both studies use the same experimental platform, a balanced dataset of realistic phishing and legitimate e-mails, and a common evaluation pipeline based on pre- and post-classification tasks. 
Study~1 focuses on how different prompting strategies shape the effectiveness of personalized LLM-generated training, while Study~2 investigates the role of personalization and duration in a $2 \times 2$ factorial design.

\begin{itemize}
    \item We provide \textbf{empirical validation} of LLMs as autonomous engines for generating phishing resilience training. Through two experiments involving 480 participants, we demonstrate that AI-generated content consistently drives significant learning gains (Accuracy, Recall, F1), establishing generative AI as a viable and scalable alternative to labor-intensive human authoring.

    \item We demonstrate that \textbf{content effectiveness is robust to prompting complexity} (Study~1). By systematically comparing four prompting strategies, we reveal that even a simple ``direct-profile'' approach performs on par with complex prompt engineering. This finding lowers the technical barrier for adoption, suggesting that non-experts can successfully deploy effective LLM-based training.

    \item We challenge the prevailing assumption that \textbf{security training requires deep psychometric personalization} (Study~2). Our results show that LLM-generated generic content is as effective as profile-based variants. This ``null result'' is a positive finding for scalability: it implies that organizations can deploy effective AI training at scale without the legal and operational costs of invasive user profiling.

    \item We identify a critical \textbf{dissociation between user perception and performance}. Our psychometric analysis reveals that while personality traits (e.g., conscientiousness) strongly predict subjective satisfaction, they do not correlate with objective learning gains. This warns practitioners against relying solely on user feedback as a metric for training success.

    \item We synthesize \textbf{evidence-based design guidelines} for next-generation \ac{PETA} programs, outlining how organizations can leverage LLMs to shift from static content libraries to dynamic, on-demand training pipelines that prioritize frequency and depth over static personalization.
\end{itemize}

\section{Background and Related Work}
Phishing is one of the most common cybersecurity attacks, as it can be carried out in various forms — such as emails, calls, and text messages — and poses a significant threat to individuals, affecting their privacy, security, and safety~\citep{apwg2024trends, ibm2024breach, ENISA2025}. As humans are the weakest link in the digital world, this attack strongly leverages human factors and a lack of training~\citep{Waddell2024HumanFactors,Desolda2022PhishingSLR}, deceiving individuals into believing that the communication they are receiving comes from a legitimate source \citep{Gallo2024HumanFactors}. For these reasons, individuals must receive proper training in order to face potential threats and be aware of the means and techniques commonly used to carry out these attacks. As technology advances and new tools are being released, \ac{PETA} interventions on phishing change accordingly. For example, this happens by creating interaction mechanisms that enable individuals to recognize elements in \acp{UI} that seem unusual \citep{Garfinkel2014UsableSecurity}. More recently, the spread of \ac{AI} and \acp{LLM} has also impacted the field of cybersecurity, thanks to their content generation capabilities, which translate into advantages in training and awareness for individuals, while also providing more resources for attackers \citep{Braun2025Simulations}. The sections below explore these topics, reporting an overview of the current state of the art.

\subsection{Phishing Training Methods}
\label{sec:training-methods}
Since phishing attacks do not exploit merely technical vulnerabilities, training individuals on their consequences is crucial for more effective prevention and detection \cite{MalloyYouGotPhished2025}. Over the past few decades, society has become increasingly aware of the importance of cybersecurity as technology has integrated into our daily lives. With this substantial shift, the integration of these themes in policies and cultures is certainly on the rise. The most common techniques for phishing prevention involve simulating emails, frontal lectures (or microlearning lessons), gamified learning sessions, or serious games \citep{Sarker2025Challenges}. They all aim to reduce individuals' susceptibility to phishing attacks, which refers to the probability or tendency of a person to respond to phishing messages, often used as an indicator of cybersecurity awareness and behavioral risks. This factor can vary significantly across demographic and organizational groups, highlighting the need for targeted training programs \citep{OnerHumanFactorsPhishing2025}. In fact, different job roles exhibit varying probabilities of clicking on phishing emails, engaging with their content, and following through with the malicious instructions they contain \citep{Bethany2025LateralPhishing}.

Serious games and gamification represent another tool for increasing motivation in learners and acquiring knowledge and skills by engaging in activities in a more playful and enjoyable manner, distinct from traditional educational approaches \citep{Katsantonis2019}. Regarding phishing, several examples in the literature exhibit different characteristics but share similar objectives. For example, "Anti-Phishing Phil" is an online game that teaches users how to spot phishing URLs \citep{Sheng2007}, and "What.Hack" is a role-playing simulation where users play through a phishing attack scenario from the attacker’s perspective \citep{Wen2019}. Similarly, gamification is a common technique used in cybersecurity education, as it integrates elements typically found in gaming environments into training and learning---e.g., points, quizzes, and competition \citep{Wen2019,Reinheimer2020}. This integration effectively mimics a competitive learning environment, which promotes greater student motivation and engagement. 

Among the most traditional approaches to education, other than typical frontal lectures, microlearning is one of the most common. They can consist of short videos (1--3 minutes), infographics, tip cards, mini-quizzes, or interactive challenges \citep{Hug2010MicroLearning}. This approach can also be more effective than game-based learning strategies \citep{Kavrestad2022ContextualLearning}.
Simulated phishing campaigns are widely used for security training, particularly in organizations and corporations. In this case, employees are sent realistic fake phishing emails via specialized domains and tools, and their responses or actions are tracked. This technique is classified as a ``\emph{learning-by-doing}'' approach, which can be implemented in various ways, for example, by conducting large-scale email campaigns \citep{Braun2025Simulations}. As researchers continue to investigate these topics, mixed approaches are also emerging; for example, when users respond to phishing emails, they should be provided training documents or small lectures in which they are taught how to recognize attacks, explaining the motivations behind why the email they responded to was malicious \citep{Cuchta2019HumanFactors}. In summary, phishing awareness training can be carried out in multiple ways; however, many users still fall for attacks even after completing training and awareness programs. Thus, the effort is shifting toward increasing the effectiveness of PETAs and personalized approaches \citep{Kavrestad2022, Reinheimer2020}.


\subsection{The Role of Personalization in Training}
Personalization plays a crucial role in ensuring that specific users are provided with \acp{UI}, messages, and information that align with their characteristics, allowing them to be trained as proficiently as possible \citep{Distler2021UsableSecurity,Kumaraguru2009Phishing,Kaufhold2025UsableSecurity}. Thus, as new technologies emerge, novel and more complex interaction paradigms can be built, allowing more effective personalization and customization (e.g., Generative \ac{AI}). 

In the context of cybersecurity, \acp{LLM} are being widely adopted for various purposes. On the one hand, they can represent a valuable tool for attackers, enabling them to increase the quantity of emails generated and improve their structure to make them more deceptive and sophisticated \citep{Qi2025SpearBot,Lopez2025Vulnerability}. In fact, \ac{LLM}-generated phishing emails can be as effective---or more---than human-generated ones, increasing the risk associated with this kind of attack \citep{Bethany2025LateralPhishing}, for example, attackers can mimic a company's or a person's communication style, including personalized information about the customers in the email \citep{Ferrag2025GenAICybersec}. On the other hand, \acp{LLM} have been exhibiting promising results in preventing and detecting phishing, by leveraging the same generative capabilities that attackers exploit to enhance their techniques \citep{Desolda2025Apollo}. This aspect makes education and training through simulations (see \cref{sec:training-methods}) more feasible and less technically challenging, thereby enabling the large-scale creation of emails and greater differentiation. An \ac{LLM}-based system has been proposed for phishing detection by analyzing emails in real-time, highlighting the importance of fine-tuning the model to obtain higher accuracy levels \citep{Meguro2025AdaPhish}. \acp{LLM} are also used to generate natural-language explanations to provide the motivations behind emails being classified as malicious for phishing. The study proposed in  \cite{Desolda2025Apollo} demonstrates that this technique increases users' trust in warnings, which results in being more understandable. Another way these tools are being employed is to generate realistic training emails tailored to the organizational context and adapt the difficulty (e.g., novice, expert), significantly reducing manual authoring effort \citep{Heiding2023LLMsDetection}. However, this topic has been only preliminarily investigated.

\subsection{Research Motivation and Objectives}
Although \acp{LLM} are increasingly applied in cybersecurity for phishing prevention and detection, a significant gap remains regarding their use in training and education, particularly in terms of personalization. The scarcity of empirical evidence on \ac{LLM}-based personalized phishing training motivated this research, with the primary objective of determining whether \acp{LLM} can effectively create \ac{PETA} programs. Specifically, we aim to investigate the impact of personalization on training effectiveness and its potential interaction with training duration.
Based on these objectives, we defined three \acp{RQ}:

\begin{enumerate}
    \item \textbf{RQ1:} Can \acp{LLM} be used to generate \ac{PETA} materials that help users improve their defense against phishing?
    \item \textbf{RQ2:} Does personalization of \ac{PETA} improve phishing detection performance compared to non-personalized training?     
    \item \textbf{RQ3:} Does training length influence the effectiveness of the intervention?
\end{enumerate}

To empirically address these questions, we designed an integrated experimental framework capable of generating, delivering, and evaluating adaptive training content. The following section outlines this approach, detailing how psychometric profiling is combined with generative AI to create personalized educational interventions.

\section{Framework Design and Approach}
Our framework operationalizes the \textbf{automated generation} of phishing training materials by leveraging the generative capabilities of \acp{LLM}. The primary goal is to streamline the production of high-quality, realistic educational content, while also providing the architecture to support psychometric profiling and personalization. The design was guided by three key principles: (i) \textbf{scalability}, enabling the rapid production of diverse training scenarios without requiring manual authoring; (ii) \textbf{human-centeredness}, ensuring that the content---whether generic or personalized---is grounded in validated psychological constructs; and (iii) \textbf{replicability}, ensuring that the workflow can be applied consistently across different experimental conditions and organizational settings.

\subsection{Designing the training activity}

The training activity was conceived as a modular, multimodal, and interactive learning experience, which combines theoretical and practical components to strengthen participants’ awareness and behavioral defenses against phishing. The design draws on evidence showing that passive consumption of textual or video materials alone rarely ensures effective learning outcomes \citep{vesin2018learning}. Additionally, it highlights that personalization, interactivity, and immediate feedback increase motivation and long-term retention \citep{Mariono_2024, csedu21}. Therefore, it includes multimodal and interactive components, such as clickable phishing examples, feedback screens, and classification tasks, to sustain attention, evoke cognitive and emotional engagement, and support metacognitive reflection \citep{akbar2024exploring}. In addition, following the logic of \textit{Kolb’s experiential learning cycle} \citep{mcleod2017kolb}, the training design proposed in this study combines moments of concrete experience, reflection, conceptualization, and active experimentation. We organized these training activities into five main modules:

\begin{enumerate}
    \item \textbf{Introduction:} This introduces users to the training course and to the problem of phishing. It provides a concise overview of the associated risks, highlights user vulnerabilities, and presents the structure and objectives of the overall training module.
    
    \item \textbf{Phishing Scenario:} This phase shows an interactive, realistic phishing example. Participants are shown a simulated phishing email incorporating standard phishing techniques (e.g., deceptive URLs, spoofed sender addresses). Suspicious elements are made interactive, allowing users to click and reveal explanations of the methods used. The scenario includes a decision point where users select their likely response and immediately receive feedback explaining the implications of their choice.
    
    \item \textbf{Defense Strategies:} In this phase, the participant is provided with actionable defense strategies against phishing, including cues for identifying legitimate messages (e.g., verifying sender domains and URLs) and reminders to secure behavioral patterns.
    
    \item \textbf{Interactive Exercises:} At this stage, participants have to complete practical exercises involving simulated phishing and legitimate emails. The emails reproduce realistic contexts and apply techniques covered in previous submodules. Users classify each message as "Phishing" or "Legitimate" and receive immediate feedback explaining the distinguishing cues. The emails in the exercises emulate those of legitimate companies, including well-known ones (e.g., Google, Microsoft, Amazon); specifically, the company names were randomized in each generated exercise.   
    
    \item \textbf{Conclusions:} The final part of the training summarizes the key takeaways about essential defense strategies and recaps best practices for identifying phishing attempts, before greeting the participant.
\end{enumerate}

\begin{figure*}[b]
    \centering
    \fcolorbox{gray}{white}{\includegraphics[width=0.9\textwidth]{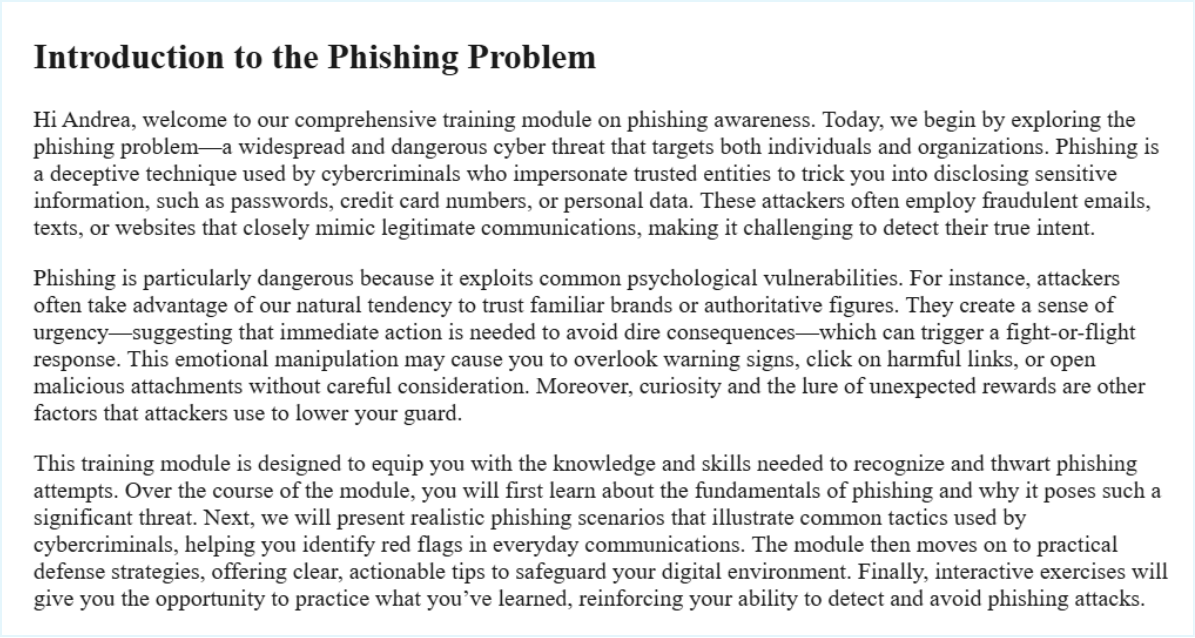}}
    \caption{Example of generated training "introduction" module.}
    \label{fig:intro}
\end{figure*}

\begin{figure*}[t]
    \centering
    \fcolorbox{gray}{white}{\includegraphics[width=0.9\textwidth]{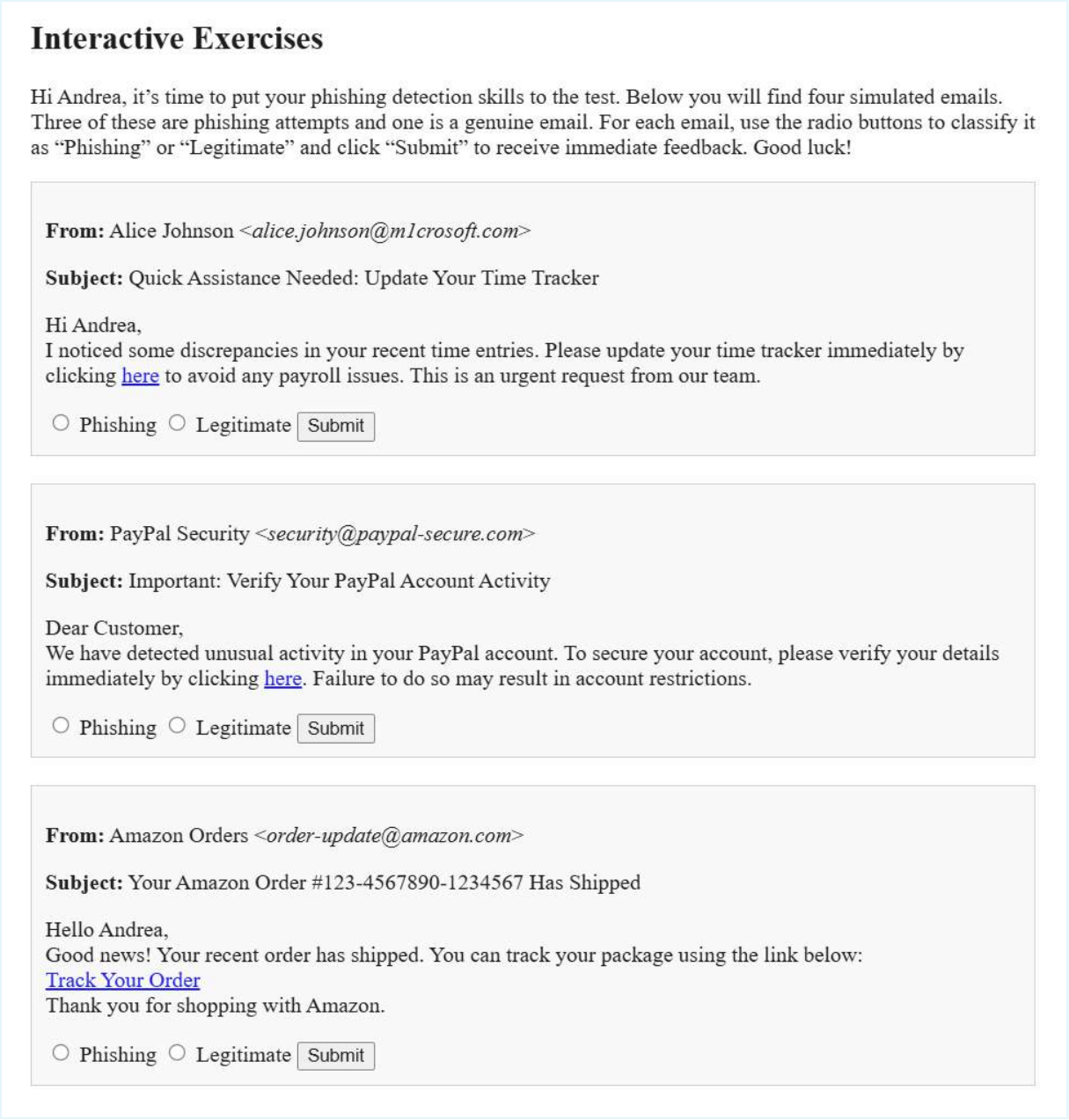}}
    \caption{Example of generated training ``exercises'' module.}
    \label{fig:exercises}
\end{figure*}

Examples of generated introduction and exercise modules are shown in Figures \ref{fig:intro} and \ref{fig:exercises}, respectively. A complete list of examples of modules generated by our framework is reported in the supplementary material.
The modular structure aims to support the progressive reinforcement of key concepts across different stages of the course, facilitating a more profound understanding through iterative exposure \citep{csedu21}. It also aligns with microlearning principles~\citep{Hug2010MicroLearning, jahnke2022microlearning}, which promote brevity, engagement, and contextual relevance. The latter is also achieved through the personalization of the training material depending on the person attending the training activity, as detailed in the following sections.

Finally, all prompts include a variable (i.e., \textit{training_length}) that allows for setting the overall duration of the training, which can be set to either "short" (9 minutes of training) or "long" (18 minutes of training). Based on this variable, each module will have a specific estimated reading time and number of words (or exercises, in the case of the fourth module). The specific lengths for each sub-module are reported in Table \ref{tab:module_lengths}. This permits training modules of varying lengths and, therefore, varying degrees of detail. 
This is a detail that we explored in Study 2 of this work.

\begin{table*}[b]
    \centering
    \caption{Module lengths varying based on the \textit{training_length} condition}
    \label{tab:module_lengths}
    \begin{tabular}{l c c}
         \hline 
         \textbf{Module} & \textbf{\textit{Short} training length} & \textbf{\textit{Long} training length} \\  
         \hline 
         Introduction & \textasciitilde1 min, \textasciitilde150 words & \textasciitilde2 min, \textasciitilde300 words  \\
         Phishing scenario & \textasciitilde2 min, \textasciitilde300 words & \textasciitilde5 min, \textasciitilde750 words \\
         Defense strategies & \textasciitilde3 min, \textasciitilde450 words & \textasciitilde6 min, \textasciitilde900 words\\
         Interactive exercises & \textasciitilde2 min, 2 exercises & \textasciitilde3 min, 3 exercises \\
         Conclusions & \textasciitilde1 min \textasciitilde150 words & \textasciitilde2 min, \textasciitilde300 words \\ \hline
         \textit{Total} & 9 minutes & 18 minutes \\
         \hline 
    \end{tabular}
\end{table*}

\subsection{Psychometric Profiling as the Basis for Adaptivity}
Personalization in our framework begins with creating a \emph{digital profile} of each user. Profiles are intended to capture and measure psychological dimensions associated with phishing susceptibility, specifically, personality traits \citep{halevi2015spear, eftimie2022spear}, susceptibility to persuasion factors \citep{abroshan2021covid}, and \ac{EI} \citep{arevalo2023human}.

To build a profile of the user, we selected three validated measurement instruments from the literature, one for each of the dimensions relevant to phishing. 
The \textit{BFI-2-XS} \citep{SOTO201769} was chosen as a concise measure of the \textit{Big Five} personality traits. Although longer inventories, such as the \textit{NEO-PI-3} \citep{McCrae01062005} or the \textit{IPIP-NEO-120} \citep{maples2014}, provide finer-grained assessments, their administration time is substantially longer. Alternative short forms such as the \textit{TIPI} \citep{GOSLING2003504} or the \textit{Mini-IPIP} \citep{Donnellan2006TheMS} are briefer but show lower internal consistency and reduced discriminant validity \citep{10.3389/fpsyg.2023.1202953, COOPER2010688}.

To assess individual differences in compliance with persuasion strategies, we employed the \textit{StP-II-B} \citep{modic2018we} scale. While other instruments such as the \textit{Persuadability Inventory} \citep{persuadability2013Bush} or the full \textit{Susceptibility to Persuasion Scale} (StP-II-B) \citep{modic2018we} include broader item sets, the \textit{StP-II-B} offers very good factorial validity \citep{TANG2024e27751} while requiring a lower administration time, making it particularly suitable for online deployment.

\ac{EI} was measured using the \textit{TEIQue-SF} \citep{petrides2009trait}, a well-validated \citep{Siegling03092015, LABORDE2016232} 30-item short form of the \textit{Trait Emotional Intelligence Questionnaire}. Alternative instruments such as the \textit{WLEIS} \citep{wong2002wong} or the \textit{Schutte Self-Report Inventory} \citep{schutte1998schutte} are also commonly employed, yet they capture narrower aspects of the construct or mix trait and ability-based components \citep{healthcare9121696, ILICETO2017274}. The \textit{TEIQue-SF}, in contrast, provides a comprehensive assessment of global trait \ac{EI} while maintaining excellent psychometric properties \citep{Cooper16082010}.
The next subsections report the details of each questionnaire used in this study.

\subsubsection{The \textit{BFI-2-XS}} 
The Big Five Inventory-2 Extra Short Form (\textit{BFI-2-XS}) \citep{SOTO201769} is a 15-item psychological instrument whose aim is to assess the five broad dimensions of personality (Agreeableness, Conscientiousness, Extraversion, Negative Emotionality, and Openness or Open-Mindedness) as defined in the Five Factor Model (FFM) \citep{McCrae1992AnIT}. It is the most compact version among the \textit{BFI-2-XS} series questionnaires, designed to reduce the administration time while preserving the robustness and clarity of the original 60-item survey. 

In the \textit{BFI-2-XS}, each personality dimension is decomposed into three subdimensions that reflect more specific behavioral and emotional tendencies of the trait. For instance, \textit{Extraversion} encompasses sociability, assertiveness, and energy level, while \textit{Conscientiousness} involves organization, productivity, and responsibility. The complete set of 15 items (one per subdimension) of the instrument is reported in the supplementary material. Respondents can select their agreement with each of the statements with a 5-point Likert scale (from \textit{1= Strongly disagree} to \textit{5= Strongly agree}); then, the scores for each of the five personality dimensions are obtained by averaging the scores of the three corresponding statements after reversing the negative items. The survey is unable to assess subdimension-level traits reliably; however, its brevity makes it optimal for large-scale surveys, especially when minimizing participants' fatigue and overall administration time is crucial.

\subsubsection{The StP-II-B} 
The \textit{Susceptibility to Persuasion Scale II (Brief)} \textit{StP-II-B} \citep{modic2018we} is a 30-item instrument designed to measure an individual’s general susceptibility to persuasive techniques, such as the psychological mechanisms that make people more likely to comply with requests, trust deceptive messages, or act impulsively when confronted with influence attempts. It is conceptually grounded in Cialdini’s principles of persuasion \citep{cialdini} along with the integration of key constructs from social psychology \citep{strathman1994consideration} and behavioral economics \citep{blais2006domain}. 

Compared to the full 54-item scale (\textit{StP-II-B}), the brief version excludes second-order constructs and retains the original 10 first-order subscales, each comprising three items. 
Respondents rate their agreement on each statement with a 7-point Likert scale (\emph{1 = ``Strongly disagree''} to \emph{7 = ``Strongly agree''}).

The subscales represent traits that attackers can exploit to increase the success rate of phishing attempts.
\textit{Premeditation}, for instance, consists in the tendency to deliberate about future consequences before acting \citep{strathman1994consideration}. Low premeditation can lead to impulsivity, which increases susceptibility to phishing and spear-phishing attacks~\cite{greitzer2021experimental, dawn2024who}. 

\emph{Consistency} reflects the motivation to maintain one's alignment between past behavior, beliefs, and public image \citep{cialdini}, while \emph{Sensation seeking} refers to a preference for excitement, novelty, and risk \citep{ARNETT1994289}, making affected individuals more susceptible to tactics that use exclusivity or novelty as scamming factors. \emph{Self-control} indicates the ability to resist impulses and regulate behavior: when it is low, individuals present high impulsivity and low inhibition. Other subdimensions are \emph{Social Influence, Similarity, Risk Preferences, Attitude toward Advertising, Need for Cognition}, and \emph{Need for Unique Choice}.
The full list of items and subdimensions of the \emph{StP-II-B} is reported in the supplementary material.

\subsubsection{The TEIQue-SF}
The \textit{Trait Emotional Intelligence Questionnaire--Short Form} (\textit{TEIQue-SF}) \citep{petrides2009trait} is a 30-item self-report inventory designed to measure the global trait of \ac{EI}, defined as “the ability to monitor one's own and others' feelings and emotions, to
discriminate among them and to use this information to guide one's thinking and actions” \citep{salovey1990emotional}. The short form was developed from the full 153-item \textit{TEIQue} by selecting two items from each of the 15 trait \ac{EI} subdimensions. Each item is rated on a 7-point Likert scale ranging from \textit{1 = Completely disagree} to \textit{7 = Completely agree}. The full questionnaire is reported in the supplementary material.

The \textit{TEIQue-SF} also allows for the computation of four broad factor scores that represent higher-order dimensions of the \ac{EI} construct, which we utilize for the purpose of personalization: \textit{Well-Being} (items 5, 9, 12, 20, 24, 27), \textit{Self-Control} (items 4, 7, 15, 19, 22, 30), \textit{Emotionality} (items 1, 2, 8, 13, 16, 17, 23, 28), and \textit{Sociability} (items 6, 10, 11, 21, 25, 26).

\subsection{Prompt Engineering and Personalization Strategies}
The core of the proposed approach is the automatic generation of training material and the integration of psychometric profiles into structured prompts for an LLM (e.g., OpenAI’s o3-mini). To this end, we explored four prompting strategies, as well as a baseline prompting method with no personalization. The employment of multiple approaches to content customization also served to mitigate the risk of suboptimal personalization that might arise from relying on a single prompting strategy. 

All of the prompting mechanisms share the same strategy for generating the training: an initial prompt given to the model with instructions to answer to subsequent requests (also called ``developer message'' in OpenAI's \textit{o} models), and 5 prompts to generate the training modules sequentially (Introduction, Phishing scenario, Defense strategies, Interactive exercises, and Conclusions). The initial prompt outlines the overall structure of the training, the requirements for the output, and content and style guidelines (e.g., content must be clear, accessible, engaging, and written in simple language, and it should refer to the user by their name). 
Moreover, if the experimental condition involves personalizing the training content, this initial prompt will also include some personalization requirements according to the 4 prompting mechanisms, defined as follows.

\begin{itemize}

    \item \textbf{Direct-profile}: we include a no-priming (zero-exemplar) condition to establish how the model performs under minimal intervention, reflecting the fact that recent work has shown \acp{LLM} can serve as decent zero-shot reasoners without any in-context examples \citep{kojima2022, brown2020}. This was achieved by directly embedding the participants' questionnaire scores into the prompt in numerical form and asking the model to tailor the training material according to the user's profile. In addition, the prompt in this case was also enriched with clear and unambiguous definitions of each trait included in the user profile, preserving consistency with the domain and preventing misinterpretation or synonymic drift by the model.
    
    \item \textbf{Few-shot priming}: In this approach, participants’ psychological profiles were again represented through numerical trait scores inside the prompt. We considered adding this priming technique since it has been shown that, when tasked with complex or variable outputs, models guided by few-shot demonstrations can outperform simpler zero-exemplar prompts \citep{SIVARAJKUMAR2024, info15110664}. 
    However, unlike the direct-profile technique, the few-shot priming condition did not include explicit definitions of each trait. Instead, the model was exposed to a limited set of illustrative examples demonstrating how content might be adapted based on the prominence of certain traits within a profile. For example: \textit{``For users high in Agreeableness or Emotionality, include emotionally resonant narratives or emphasize social impact.''} or \textit{``For users more influenced by social proof or authority, highlight how attackers might exploit these principles.''}
    The objective of this setup is to convey a general understanding of how personalization can be achieved, rather than prescribing precise adjustment rules for each trait included in the profile. 
    
    \item \textbf{Table-based priming}: In this condition, the participant’s psychological profile was once more represented through numerical trait scores. This condition was explored because previous research suggests that the ability of LLMs to interpret and act on structured input is strongly dependent on their format \citep{sui2024}, and that structured prompts can improve reliability, control, and alignment of models' outputs \citep{CHEN2025101260}.
    However, unlike the few-shot priming approach, the model was not expected to infer general personalization strategies from a small number of examples. Instead, it was provided with an explicit and structured table outlining concrete content adaptation strategies for each trait. The table consisted of three columns: the first lists all psychological traits included in the profile, while the second and third specify how content should be adapted for individuals exhibiting, respectively, high or low values on each corresponding trait. 
    
    \item \textbf{Guideline-based priming:} Drawing on evidence that task decomposition into subrules---a process also known as \emph{itemization}---improves performance \citep{mishra-etal-2022-reframing, vilakati2025}, and that context length negatively impacts efficiency \citep{wan2024efficientlargelanguagemodels}, we employed a final technique that does not require including the participant’s profile inside the prompt. Instead, we determine the three most salient traits by normalizing all profile values within the $(0,1)$ range and selecting the three most extreme scores---in other words, those closest to 0 or 1---which represent the strongest tendencies in either direction. 
    Each trait is associated with specific customization guidelines, which were derived from a predefined repository containing structured adaptation principles for both high and low trait values (reported in the Additional Material). Each guideline was organized into three components: (1) \textit{Communication Style}, describing how to frame the tone and emotional register of the message; (2) \textit{Learning Content}, specifying the educational approach and depth of information; and (3) \textit{Phishing Scenario}, detailing the type and framing of the illustrative examples to be used.
    For instance, for the trait \textit{Negative Emotionality}, the \textit{Communication Style} directive for high values recommended ``reassuring and supportive messaging focused on empowerment rather than fear or alarmist language'', whereas for low values it prescribed a ``straightforward and matter-of-fact tone providing comprehensive insights into risks and consequences''. The learning content and the communication style are presented as a set of personalization requirements in the initial prompt, in place of the complete user profile. 
\end{itemize}

All the personalization guidelines used to tune the LLM in the content generation phase have been developed and double-checked by two distinct researchers, experts in Human-Centered Cyber Security, taking into account the characteristics of the different traits. As previously mentioned, this design not only allowed us to contrast personalization with non-personalization but also to investigate which prompting strategy yields the most effective training materials. For each training strategy, all the prompts are reported in the supplementary material.

\begin{figure*}[t!]
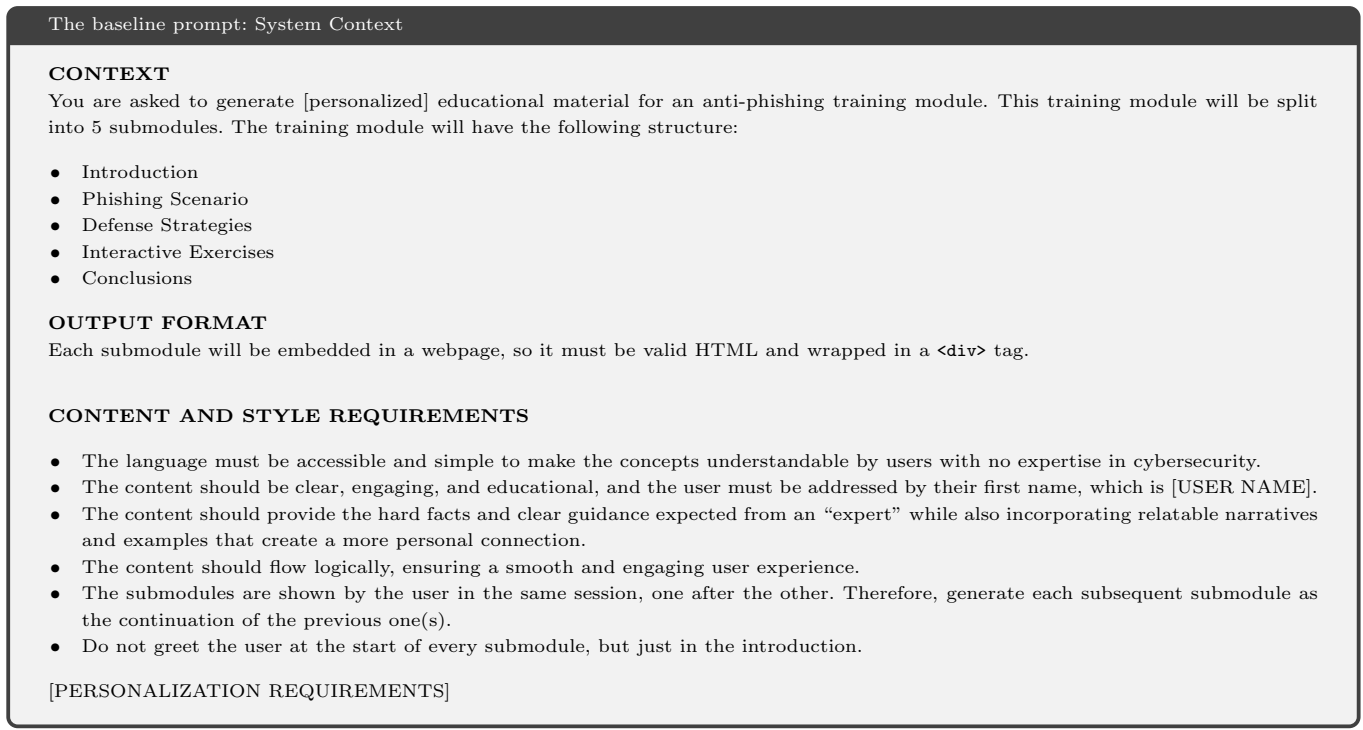
 
    \centering
    \small 
    
    \begin{tcolorbox}[colback=llgray, title=The baseline prompt: System Context]
    
    \textbf{CONTEXT}
    
    You are asked to generate [personalized] educational material for an anti‑phishing training module. This training module will be split into 5 submodules. The training module will have the following structure:
    
    \begin{itemize}
        \item  Introduction
        \item  Phishing Scenario
        \item  Defense Strategies
        \item  Interactive Exercises
        \item  Conclusions
    \end{itemize}
    
    \textbf{OUTPUT FORMAT}
    
    Each submodule will be embedded in a webpage, so it must be valid HTML and wrapped in a \texttt{<div>} tag.\\[.5\baselineskip]
    
    \textbf{CONTENT AND STYLE REQUIREMENTS}
    
    \begin{itemize}
        \item  The language must be accessible and simple to make the concepts understandable by users with no expertise in cybersecurity.
        \item  The content should be clear, engaging, and educational, and the user must be addressed by their first name, which is [USER NAME].
        \item  The content should provide the hard facts and clear guidance expected from an “expert” while also incorporating relatable narratives and examples that create a more personal connection.
        \item  The content should flow logically, ensuring a smooth and engaging user experience.
        \item  The submodules are shown by the user in the same session, one after the other. Therefore, generate each subsequent submodule as the continuation of the previous one(s).
        \item  Do not greet the user at the start of every submodule, but just in the introduction.
    \end{itemize}
    
    [PERSONALIZATION REQUIREMENTS]
    \end{tcolorbox}
    \caption{Complete structure of the baseline prompt including context, format, and style constraints.}
    \label{fig:baseline_prompt}
\end{figure*}

Each module description also includes the approximate target duration, depending on the training length (as specified in Table \ref{tab:module_lengths}). It is worth mentioning that, for the guideline-based prompting condition only, the module description of the Phishing scenario specifies the personalization guidelines.  

\section{Study 1 - Identify the best prompting strategy}
Before investigating the best training strategies in the large (e.g., with or without personalization and training duration) , we conducted an initial study to explore the benefits of four prompting techniques on the knowledge acquired by participants through the related training. 

\subsection{Participants}
A total of 80 participants took part in the study. They were recruited through the Prolific platform, which offers access to diverse and pre-screened populations. We established eligibility criteria to ensure comparability across conditions: participants needed to be fluent in English, reside in Western Europe, use a laptop or desktop computer, and possess at least basic digital literacy skills. These constraints minimized confounds associated with language comprehension or device capabilities, while still maintaining a heterogeneous and realistic user base. The final sample consisted of 41 men, 39 women (with no non-binary individuals), with an average age of 35.33 years ($SD = 10.97$). 

The participation in the study lasted approximately 20 minutes, and participants were rewarded with \pounds3.00, in line with Prolific's recommended participation fee of \pounds9.00/hour. 

\subsection{Experimental Design}
The experiment employed a 4 (Prompting Technique: direct-profile, Few-shot priming, Table-based priming, Guideline-based priming) × 2 (Phase: Pre vs. Post) mixed design, with between-subjects and within-subjects factors. Each participant was exposed to a single prompting condition but completed both pre- and post-training classification tasks. This design enabled us to examine not only overall differences between prompting strategies but also how participants’ phishing detection performance changed as a result of the training intervention. 

The between-subjects factor (\textit{Prompting Technique}) allowed comparison of the four LLM prompting methods used to generate the training material, while the within-subjects factor (\textit{Phase}) captured individual learning gains over time. The dependent variables were behavioral performance metrics---accuracy, recall, and F1-score---computed separately for the pre- and post-test phases. Together, this configuration enabled us to test whether specific prompting strategies yielded larger improvements in participants’ ability to correctly classify phishing emails versus legitimate ones.

The analyses for Study~1 were guided by four hypotheses concerning the effectiveness of LLM-generated phishing-awareness training and the role of prompting strategies:

\begin{itemize}
    \item \textbf{H1 (Training effectiveness for phishing resilience).}  AI-generated training will lead to an improvement in resilience against phishing attacks, irrespective of the prompting strategy. In other words, participants will exhibit significant improvements from Pre to Post across all performance metrics (Accuracy, Recall, F1).

    \item \textbf{H2 (Prompting differences).}  
    The magnitude of improvement ($\Delta = \text{Post} - \text{Pre}$) will differ across prompting conditions (Few-shot, Table-based, Guideline-based, Direct-profile).

    \item \textbf{H3 (Subjective reactions).}  
    Prompting strategies will influence participants’ subjective evaluations of the training (interest, involvement, usefulness, trainer satisfaction, expectations).

    \item \textbf{H4 (Best-performing strategy).}  
    One or more prompting variants will emerge as particularly effective based on the combined pattern of high Post scores and substantial improvement.
\end{itemize}

\subsection{Material}


To enhance the quality and generalizability of our study, we constructed a new dataset of 36 emails designed to balance message genuineness (18 phishing vs.\ 18 legitimate), topics, and classification difficulty. Drawing on recent analyses of phishing trends---such as the \textit{APWG Phishing Activity Trends Report}\footnote{https://docs.apwg.org/reports/apwg_trends_report_q2_2024.pdf} and the \textit{IBM Data Breach Report}~\cite{ibm2024breach}---we identified six representative email topics: Suspicious Activity, Payment Required, Failed Login Attempt, Gift Card, Action Required, and Tracking Information.
To simulate realistic communication patterns, senders were selected from well-known brands and online platforms, including Facebook, Nike, PayPal, Shein, and Amazon, for a total of 24 senders. Among these, 12 senders had corresponding genuine and phishing versions, while another 12 senders were distinct, comprising 6 phishing emails and 6 legitimate emails.

\begin{figure*}[t]
    \centering
    \begin{subfigure}{0.75\linewidth}
        \centering
        \fcolorbox{gray}{white}{\includegraphics[width=\linewidth]{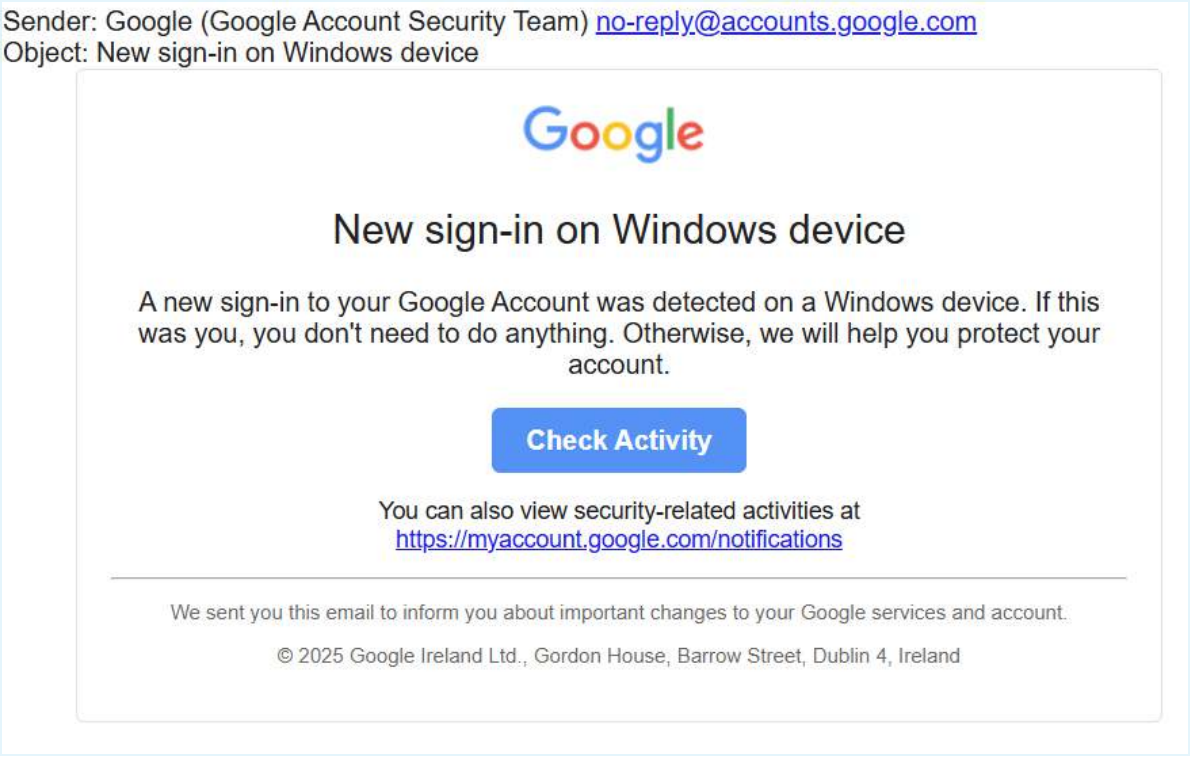}}
        \caption{Legitimate email}
    \end{subfigure}
    \vspace{0.5em}
    \begin{subfigure}{0.75\linewidth}
        \centering
        \fcolorbox{gray}{white}{\includegraphics[width=\linewidth]{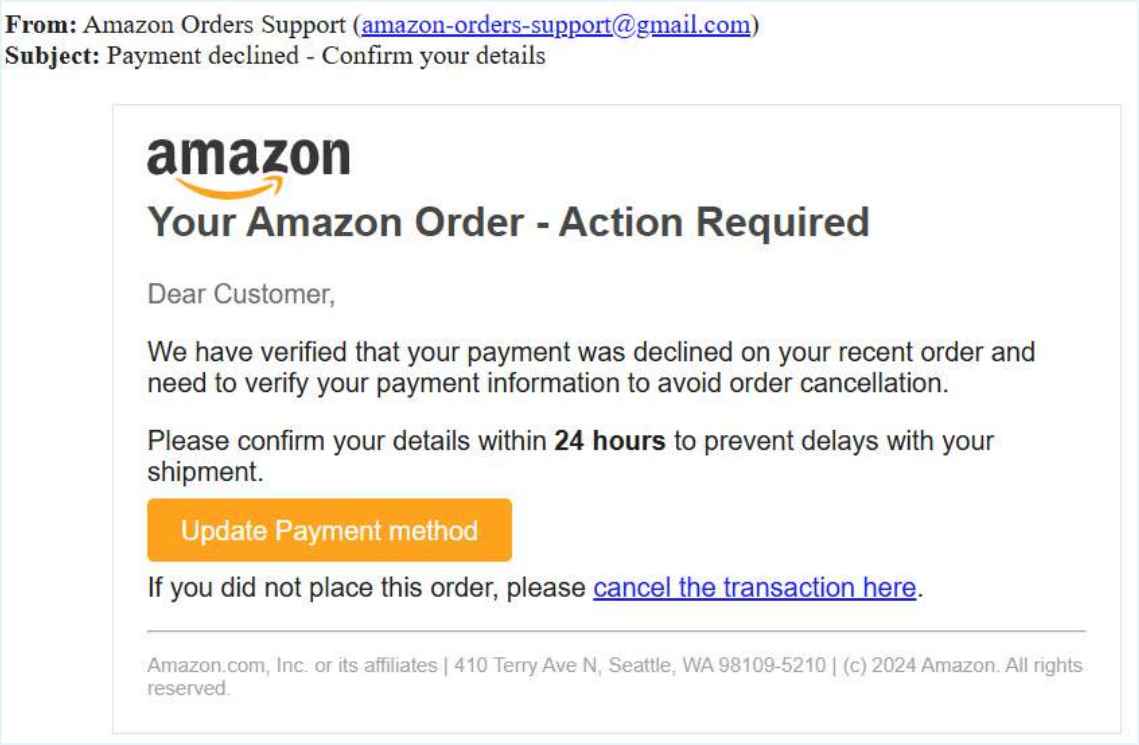}}
        \caption{Phishing email}
    \end{subfigure}
    \caption{Examples of email samples used in the study: (a) legitimate email and (b) phishing email.}
    \label{fig:email_examples}
\end{figure*}

Beyond topic and type, we also balanced the intended difficulty of detection across three levels (easy, medium, hard). To operationalize this, we applied the first component of the \textit{United States' National Institute of Standards and Technology (NIST) Phish Scale}~\citep{1220571}, which assesses the inherent characteristics of an email by quantifying its observable cues of suspiciousness. This component categorizes cues into five types---errors, technical indicators, visual presentation indicators, language and content cues, and common social-engineering tactics---and requires counting each instance of all 23 defined cues. The resulting total count is then mapped to one of three difficulty-related cue categories: few cues (1--8), some cues (9--14), or many cues (15+). In the case of phishing messages, fewer cues indicate that emails are inherently more challenging for users to detect, while legitimate emails are harder to identify if many suspicious cues are present.
The second component of the \textit{NIST Phish Scale} assesses premise alignment, and was not applicable in our case, as it requires detailed a priori knowledge of the target audience’s roles, contextual practices, and prior exposure to phishing training. 
Two samples of the dataset are reported in Figure \ref{fig:email_examples}, showing both a legitimate and a phishing email. The complete dataset of 36 emails is available in the supplementary material.

In addition to the dataset, an ad-hoc questionnaire has been designed to gain a deeper understanding of the participants' impressions and experiences during the completion of the training program. The details of the questions are reported in Table \ref{tab:training_reaction_items}.

\begin{table}[ht]
    \centering
    \caption{Training reaction questionnaire items.}
    \label{tab:training_reaction_items}
    \begin{tabular}{@{}cp{0.8\linewidth}@{}}
    \toprule
    \#     &  Item\\\midrule
    1     &  ``How interesting did you find the course?''\\
    2     &  ``How involved did you feel during the course activities?''\\
    3     &  ``How much do you think the course has improved your skills?''\\
    4     &  ``How useful do you think the course will be for your daily work?''\\
    5     &  ``How satisfied are you with the trainer and his ability to convey the content?''\\
    6     &  ``How satisfied do you feel that the course met your expectations?''\\
    \bottomrule
    \end{tabular}
\end{table}

We have finally developed a dedicated online platform, utilizing the Laravel framework, which enables the delivery of study materials, the secure collection of participant responses, and the consistent management of all interaction logs within a single, integrated infrastructure.

\subsection{Procedure}

The entire study procedure was approved by the Research Ethics Committee of the University of Bari.

After the Prolific task was accepted, the participants were redirected to the web platform where the study was hosted. According to the ethical guidelines, participants were asked to provide their informed digital consent. All data on participants and interactions were completely anonymized and securely stored on the university servers. Any concerns that the participants may have had could be addressed after the study. Once consent was obtained, all participants were randomly assigned to one of the four experimental conditions, with balanced participation across conditions. The study then unfolded in five phases, completely guided by the platform. 

\begin{figure*}[ht]
    \centering
    \includegraphics[width=0.6\linewidth]{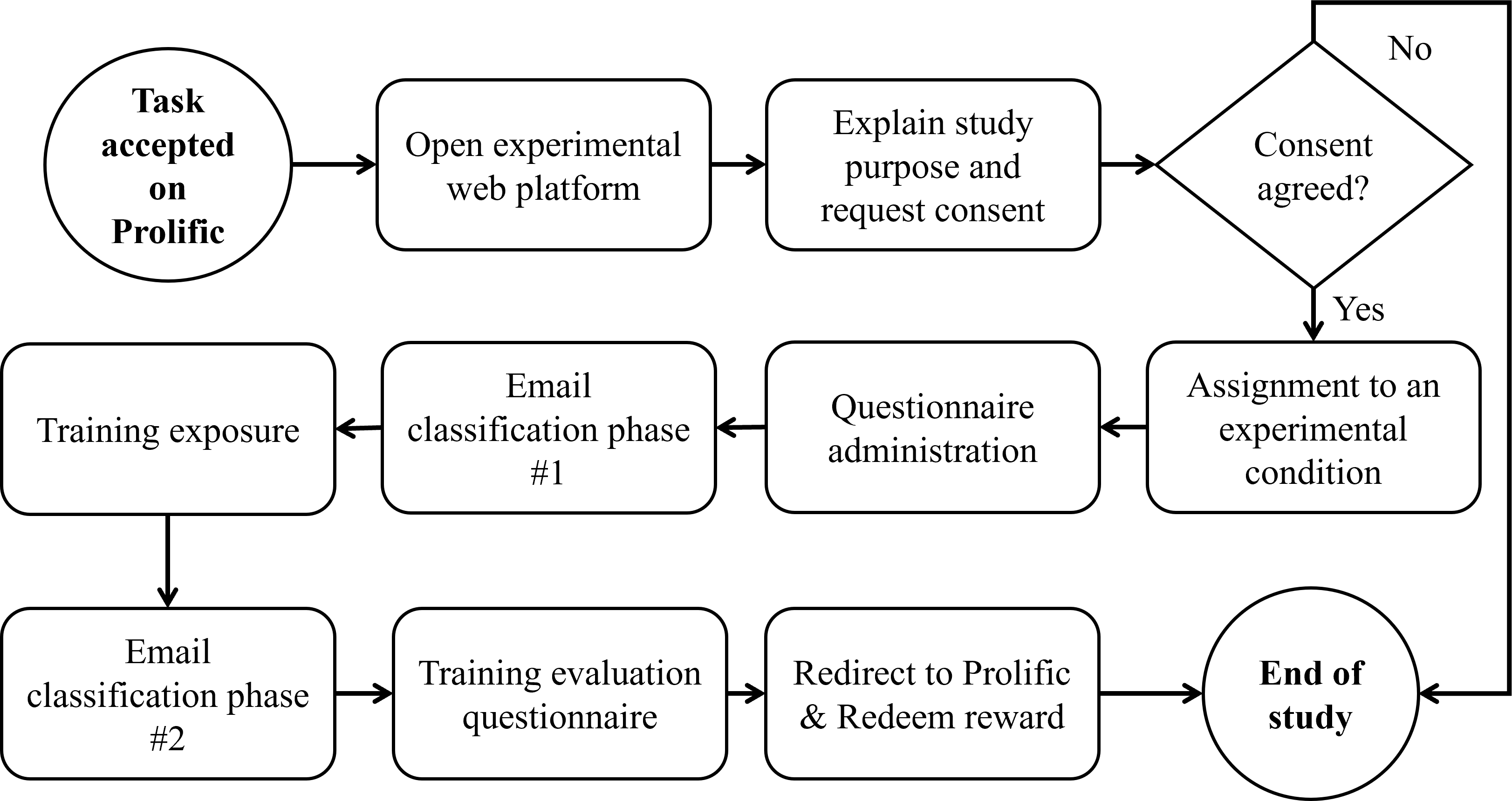}
    \caption{Study procedure for the evaluation of \ac{LLM}-generated trainings study}
    \label{fig:training_study_procedure}
\end{figure*}

In the first phase, participants completed three validated psychometric questionnaires: the BFI-2-XS for personality traits, the StP-II-B for persuasion susceptibility, and the TEIQue-SF for trait \ac{EI}. 

In the second phase, participants completed a pre-test classification task consisting of 12 e-mails selected from the dataset using a strict stratified sampling procedure. Each participant was randomly exposed to 6 legitimate and 6 phishing e-mails; within each class, the messages included 2 easy, 2 medium, and 2 hard items, and each of the six topics was represented exactly once. For each e-mail, participants indicated whether they believed it was phishing or legitimate. 

The third phase consisted of the training intervention. All trainings were generated by the \ac{LLM} (OpenAI's \textit{o3-mini with medium reasoning}) using the same modular structure (Introduction, Scenario, Defense Strategies, Exercises, Conclusion) and were personalized based on each participant’s psychometric profile. This ensured consistency in format while allowing the content to adapt to individual user characteristics.

In the fourth phase, participants completed the post-test classification task, which followed the same stratified sampling procedure as the pre-test: 12 e-mails (6 legitimate, 6 phishing), balanced across difficulty levels and topics. Thanks to the structure of the 36-item dataset and the sampling algorithm, the post-test set was always non-overlapping with the pre-test set while still satisfying all stratification constraints. This allowed us to assess learning effects using two structurally equivalent but completely distinct sets of messages.

Finally, in the fifth phase, participants completed the post-study questionnaire. This combination of quantitative and qualitative feedback offered a richer picture of user experience than performance metrics alone.

\subsection{Measures}
Our evaluation combined objective and subjective measures. Behavioral performance was captured through accuracy, recall, and F1-score, with all metrics computed for both the pre- and post-test phases. 

Subjective measures complemented this picture by focusing on the user experience of training. The reaction questionnaire gauged perceived engagement, satisfaction with the trainer (as conveyed by the LLM-generated content), the perceived utility of the training in daily work, and the extent to which it met expectations.

\subsection{Results}
This section presents the results of Study~1, organized around the four hypotheses concerning the effects of different \emph{prompting strategies} in LLM-generated phishing-awareness training. We first summarize descriptive patterns and the overall training effect, and then address each hypothesis in turn. 

\subsubsection{Statistical Analysis}
To assess the effectiveness of the training and compare the prompting strategies (H1--H4), we employed a mixed-design Analysis of Variance (ANOVA) with \textit{Phase} (Pre vs. Post) as the within-subjects factor and \textit{Prompting Technique} (4 levels) as the between-subjects factor.
Additionally, to directly compare the magnitude of learning gains across conditions, we performed a one-way ANOVA on the improvement scores ($\Delta = \text{Post} - \text{Pre}$).
To ensure robust inference in pairwise comparisons and within-condition tests, we applied the \textbf{Holm--Bonferroni} correction to control the Family-Wise Error Rate (FWER).

\subsubsection{Preliminary descriptive patterns}
Table~\ref{tab:condition_summary} reports the pre- and post-means for each prompting condition. 
Across all four prompting strategies, performance improved from Pre to Post on all three dependent variables (Accuracy, Recall, F1). 
In line with our expectations, the largest gains tended to emerge for \emph{Recall}---particularly in the \emph{Table-based} and \emph{direct profile} conditions---suggesting that the training primarily enhanced participants’ ability to detect phishing cues. 
Improvements in \emph{Accuracy} and \emph{F1} were also present, though generally smaller in magnitude. Figure~\ref{fig:study1_prepost_box} complements these summaries by showing the full distribution of Pre--Post scores for all three metrics, aggregated across prompting conditions. The boxplots highlight a systematic rightward shift from Pre to Post, with particularly pronounced changes for Recall.

\begin{table*}[t]
\centering
\caption{Summary of Pre and Post performance by prompting condition. Values are means; $\Delta$ denotes $\text{Post} - \text{Pre}$.}
\label{tab:condition_summary}
\begin{tabular}{@{}lcccccccccc@{}}
\toprule
 & & \multicolumn{3}{c}{Accuracy} & \multicolumn{3}{c}{Recall} & \multicolumn{3}{c}{F1} \\
\cmidrule(lr){3-5} \cmidrule(lr){6-8} \cmidrule(lr){9-11}
Condition & $N$ & Pre & Post & $\Delta$ & Pre & Post & $\Delta$ & Pre & Post & $\Delta$ \\
\midrule
Few-shot        & 20 & 0.737 & 0.800 & 0.063 & 0.808 & 0.850 & 0.042 & 0.761 & 0.807 & 0.046 \\
Table-based     & 20 & 0.713 & 0.797 & 0.083 & 0.750 & 0.875 & 0.125 & 0.729 & 0.802 & 0.073 \\
Guideline-based & 20 & 0.742 & 0.805 & 0.063 & 0.783 & 0.883 & 0.100 & 0.746 & 0.817 & 0.071 \\
Direct profile      & 20 & 0.723 & 0.822 & 0.098 & 0.742 & 0.908 & 0.167 & 0.727 & 0.833 & 0.105 \\
\bottomrule
\end{tabular}
\end{table*}

\begin{figure*}[t]
  \centering
  \includegraphics[width=\linewidth]{figures/fig_boxplot_prepost_allmetrics.png}
  \caption{Study~1. Pre--Post performance distributions for Accuracy, Recall, and F1-score, aggregated across prompting conditions. Each panel shows boxplots for the Pre and Post phases, illustrating consistent improvements across all three metrics.}
  \label{fig:study1_prepost_box}
\end{figure*}

Among the four variants, the \emph{direct profile} condition yielded the strongest Post-test performance, with the highest values for Accuracy ($M = .822$), Recall ($M = .908$), and F1 ($M = .833$). The three structured prompting strategies (Guideline-based, Table-based, Few-shot) also produced clear improvements, with Post-test Recall ranging from {.850} to {.883} and Post-test F1 ranging from {.802} to {.817}. Although the structured variants achieved slightly lower absolute Post means than \emph{direct profile}, their trajectories reflected consistent learning gains across all metrics.

Overall, these descriptive trends highlight two key observations: (i) all LLM-generated trainings led to measurable performance improvements, and (ii) the \emph{direct profile} strategy produced the strongest outcomes in absolute terms.

\subsubsection{H1: Does AI-generated training improve phishing resilience?}

Hypothesis~1 predicted significant improvements from Pre to Post across conditions.  
A mixed ANOVA with \textit{Phase} (Pre vs.\ Post) as a within-subjects factor and \textit{Prompting} as a between-subjects factor confirmed a strong and significant main effect of Phase for all three metrics (Table~\ref{tab:anova_mixed_study1}), i.e., \emph{Accuracy} ($p=.001$), \emph{Recall} ($p<.001$), \emph{F1} ($p=.002$).

\begin{table*}[b]
\centering
\caption{Summary of mixed ANOVA with \textit{Phase} (Pre vs.\ Post) as within-subject factor and \textit{Prompting} (4-level) as between-subjects factor, for each dependent variable.}
\label{tab:anova_mixed_study1}
\begin{tabular}{@{}llccccl@{}}
\toprule
DV & Effect & $F$ & $DF_1$ & $DF_2$ & $p$ & $\eta^2$ \\
\midrule
Accuracy & Prompting                  & 0.11 & 3 & 76 & .957  & .004 \\
Accuracy & Phase                      & 11.62 & 1 & 76 & .001  & .133 \\
Accuracy & Phase $\times$ Prompting   & 0.14 & 3 & 76 & .935  & .006 \\
\midrule
Recall   & Prompting                  & 0.10 & 3 & 76 & .960  & .004 \\
Recall   & Phase                      & 15.83 & 1 & 76 & <.001 & .172 \\
Recall   & Phase $\times$ Prompting   & 0.92 & 3 & 76 & .435  & .035 \\
\midrule
F1       & Prompting                  & 0.10 & 3 & 76 & .959  & .004 \\
F1       & Phase                      & 10.04 & 1 & 76 & .002  & .117 \\
F1       & Phase $\times$ Prompting   & 0.27 & 3 & 76 & .843  & .011 \\
\bottomrule
\end{tabular}
\end{table*}

Within-condition tests of improvement (detailed in the supplementary material) further support this conclusion. 
All four prompting strategies produced positive $\Delta$ values, and two conditions showed statistically significant or marginally significant improvements after Holm correction: \emph{direct profile} yielded significant or borderline-significant gains across all three metrics (Accuracy, Recall, F1), and the \emph{Table-based} strategy showed a significant improvement in Recall. Effect sizes ranged from small to moderate.

These results offer strong support for \textbf{H1}:  
\emph{LLM-generated training reliably improves phishing detection performance, particularly Recall.}

\subsubsection{H2: Do prompting strategies differ in effectiveness?}

\begin{figure*}[t]
  \centering
  \includegraphics[width=\linewidth]{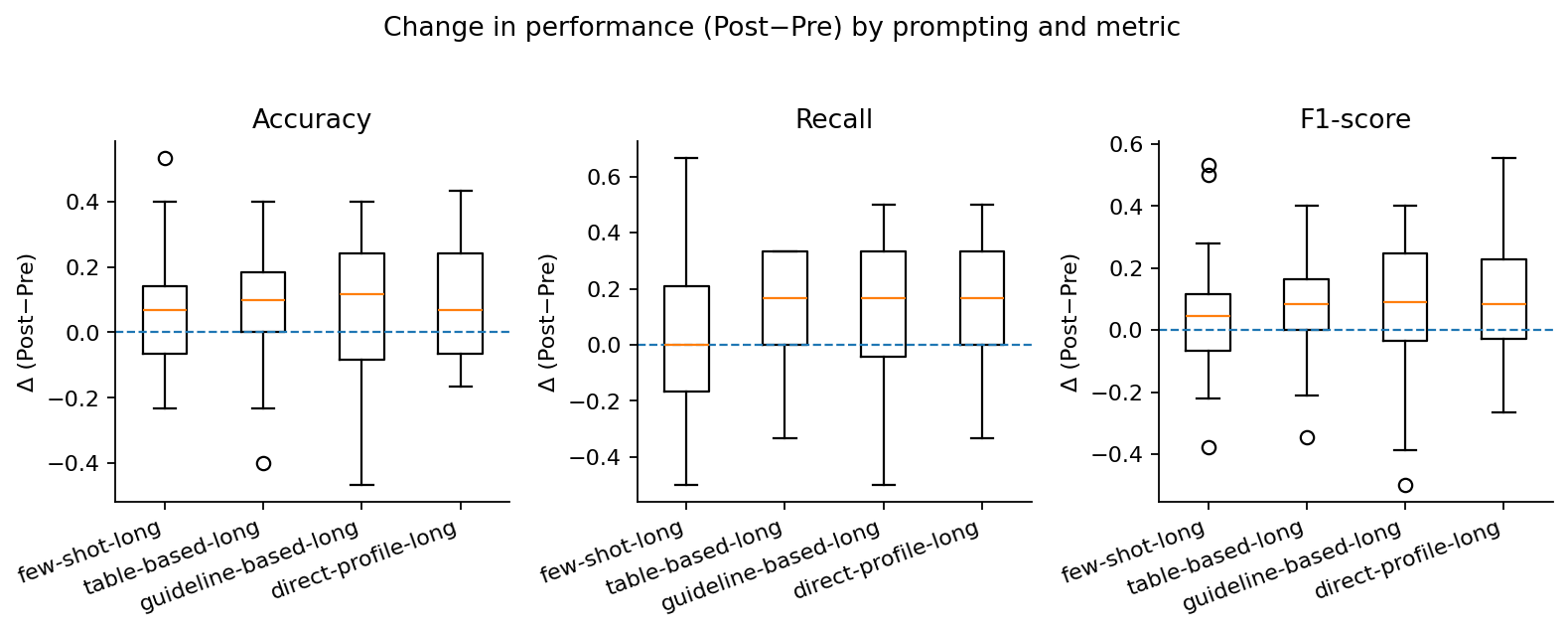}
  \caption{Study~1. Change scores ($\Delta = \text{Post} - \text{Pre}$) for Accuracy, Recall, and F1-score across prompting conditions. Each panel displays the distribution of improvements for a single metric. All conditions exhibit positive gains, and the overlap between boxplots reflects the absence of reliable differences between prompting strategies.}
  \label{fig:study1_delta_box}
\end{figure*}

Hypothesis~2 predicted differences across prompting variants in the magnitude of improvement.  
A one-way ANOVA on $\Delta$ scores (Table~\ref{tab:anova_delta}) did not reveal 
significant differences for any DV, i.e., {Accuracy} ($p=.935$), \emph{Recall} ($p=.435$), \emph{F1} ($p=.843$). Figure~\ref{fig:study1_delta_box} provides a complementary view of these results by visualizing the distribution of change scores ($\Delta = \text{Post} - \text{Pre}$) for each prompting condition and metric. The boxplots show consistently positive improvements across all conditions, with highly overlapping distributions and no prompting variant standing out as clearly superior.

\begin{table}[t]
\centering
\caption{One-way ANOVAs on improvement scores ($\Delta = \text{Post} - \text{Pre}$) across prompting conditions for each dependent variable (Study~1).}
\label{tab:anova_delta}
\begin{tabular}{@{}lccc@{}}
\toprule
DV & $F(3,76)$ & $p$ & $\eta^2$ \\
\midrule
Accuracy & 0.142 & .935 & .006 \\
Recall   & 0.921 & .435 & .035 \\
F1       & 0.275 & .843 & .011 \\
\bottomrule
\end{tabular}
\end{table}

Thus, \textbf{H2 is not supported}:  
\emph{all prompting variants produced similar improvements in performance.}

\subsubsection{H3: Do prompting strategies influence subjective reactions?}

\begin{figure*}[t]
  \centering
  \includegraphics[width=1\linewidth]{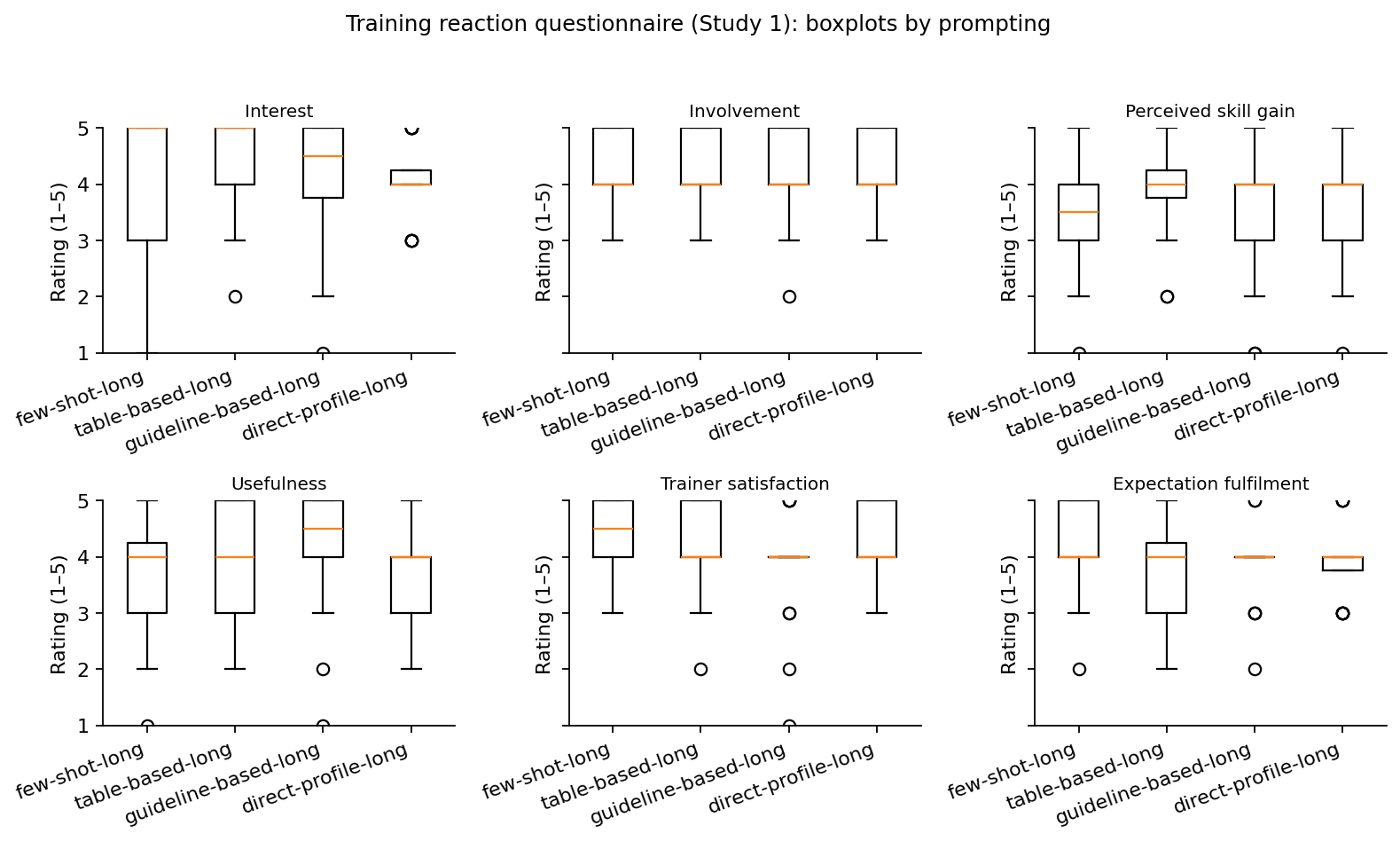}
  \caption{Study~1. Boxplots of training reaction ratings (Q1--Q6) across prompting conditions. 
  Each panel corresponds to one questionnaire item, with higher values indicating more positive evaluations.}
  \label{fig:study1_qitems_boxplots}
\end{figure*}

Hypothesis~3 posited differences in subjective evaluations across prompting conditions. Figure \ref{fig:study1_qitems_boxplots} shows box plots of subjective training reactions (Q1--Q6) across prompting conditions, illustrating uniformly high ratings and minimal differences between variants. However, one-way ANOVAs for each item revealed no significant main effect of prompting (all $p > .10$; e.g., Q5: $F(3,76)=1.38$, $p=.255$). 

Therefore, \textbf{H3 is not supported}:  
\emph{participants evaluated the training positively and similarly across prompting variants}.

\subsubsection{H4: Which prompting variant performs best overall?}

Although conditions did not differ significantly in improvement scores (H2), a comparison of 
final post means and effect sizes indicate a clear pattern.  
The \emph{direct profile} condition achieved: the highest Post accuracy (.822), the highest Post recall (.908), the highest Post F1 (.833), and the largest or near-largest $\Delta$ values across all metrics.

Structured prompting strategies (Guideline-based, Table-based, Few-shot) performed closely and consistently, 
showing robust but slightly smaller gains. Thus, \textbf{H4 is supported at the descriptive and practical level}:  
\emph{direct profile emerges as the most effective prompting variant in Study~1}, although differences are not statistically significant.

\subsubsection{Summary}

To sum up, (i) all prompting variants produced positive and statistically reliable Pre--Post improvements across key phishing detection metrics, as captured by the main effect of \textit{Phase}; (ii) one-way ANOVAs on $\Delta$ found no reliable differences in the magnitude of improvement between prompting strategies, with very small associated effect sizes; and (iii) descriptive comparisons indicate that the 
\emph{direct profile} condition emerges as the most effective overall. It achieved the highest absolute post-training scores across Accuracy, Recall, and F1, together with the largest Pre--Post improvements on all three metrics.

Among the structured prompting strategies, the \emph{Table-based} variant represents the strongest competitor: although its final performance remained slightly below that of \emph{direct profile}, it produced substantial learning-related gains---particularly in Recall and F1---and consistently outperformed the other structured alternatives.

Importantly, although these differences did not reach statistical significance, the overall pattern is consistent: (i) direct profile shows a descriptive advantage, (ii) its performance is stable across all three metrics, and (iii) there is no evidence supporting more elaborate prompting strategies. Taken together, these observations suggest that direct profile is both effective and highly parsimonious. Unlike Few-shot, Table-based, or Guideline-based prompting, the \emph{direct profile} variant requires no additional prompt engineering effort, no curated examples, and no pre-structured templates. This makes it not only the most effective option descriptively, but also the most parsimonious and 
easiest to standardize in large-scale or automated training pipelines. 

It is important to note that Study 1 was designed as a selection study rather than a definitive confirmatory trial. While the sample size ($N=20$ per cell) limited the statistical power to detect small effect sizes between prompting strategies, the consistent descriptive superiority of the direct profile condition provided a sufficient signal for selection. Furthermore, we prioritized \emph{parsimony}: since complex prompting strategies (e.g., Few-shot, Table-based) did not yield statistically significant advantages over the simpler direct profile approach, the latter was identified as the most efficient implementation.

For these reasons, \textbf{direct profile was selected as the prompting strategy for the follow-up experiment (Study~2)}, enabling us to focus on the two factors most likely to influence training effectiveness---\textit{personalization} and \textit{training length}---without 
introducing unnecessary complexity or design overhead from more elaborate prompting schemes.

\section{Study 2 - Identify the best training strategies}

We designed a second controlled experiment to investigate the effectiveness of phishing training generated by LLMs based on two key features: the length and personalization of the content. 

\subsection{Participants}
A total of 400 participants took part in this second study. Similar to Study 1, participants were recruited through the Prolific platform using the same eligibility criteria. The final sample consisted of 204 men and 196 women (with no non-binary individuals), with an average age of 37.59 years ($SD = 13.03$). This diversity provided a strong basis for examining the robustness of our findings.


\subsection{Experimental Design}
Building on the findings of Study 1, this follow-up experiment employed a $2 \times 2 \times 2$ mixed design. Two between-subjects factors were manipulated: \textit{Personalization} (personalized vs.\ generic) and \textit{Training length} (short vs.\ long), and one within-subjects factor captured performance over time (\textit{Phase}: Pre vs.\ Post). Thus, each participant was randomly assigned to one of four training conditions:

\begin{enumerate}
    \item \textbf{Generic short}: Non-personalized training (9 minutes). The LLM generated a generic phishing-awareness module with no reference to the participant's profile.
    \item \textbf{Generic long}: Non-personalized training (18 minutes). Same content-generation strategy as the generic short condition, but with an extended duration.
    \item \textbf{Personalized short}: Personalized training (9 minutes). The LLM-generated content is tailored to each participant’s profile using the \emph{direct profile} personalization procedure adopted in Study~1.
    \item \textbf{Personalized long}: Personalized training (18 minutes). The same personalization strategy applies, with additional depth and examples enabled by the longer duration.
\end{enumerate}

Study 2 was thus designed to determine whether the two key properties of LLM-generated phishing training — \textit{personalization} and \textit{training length} — affect users' ability to detect phishing emails after the intervention. Based on the results of Study~1 and prior work on adaptive training and security behavior change, we formulated the following hypotheses.

\textit{H1 (Training effectiveness for phishing resilience)}. AI-generated training will lead to an improvement in resilience against phishing attacks, irrespective of the prompting strategy. In other words, participants will exhibit significant improvements from Pre to Post across all performance metrics (Accuracy, Recall, F1). Given the higher number of participants, this hypothesis aims to confirm or refute the preliminary results of H1.

\textit{H2 (Personalization effect).}
Participants receiving \emph{personalized} training are expected to achieve higher post-training performance than those receiving \emph{generic} training. 
This prediction follows from both the superior performance of the personalized \emph{direct profile} condition in Study~1 and prior evidence that personalized security interventions can enhance relevance, engagement, and retention.

\textit{H3 (Length effect).}
Participants exposed to \emph{long} training are expected to outperform those receiving \emph{short} training, as extended interventions allow for more examples, richer explanations, and deeper processing of phishing cues.

\textit{H4 (Interaction).}
We further hypothesize a positive interaction between \textit{personalization} and \textit{length}. In particular, longer training may amplify the benefits of personalization, enabling the LLM-generated tailored content to provide more targeted examples and clarifications aligned with the participant’s profile.

\textit{H5 (Subjective experience).}
Personalization and longer training are also expected to positively influence users’ reactions to the training (e.g., perceived usefulness, engagement, and satisfaction), reflecting higher perceived relevance and instructional quality.

\textit{H6 (Psychometric predictors of performance and training reactions).}
Individual differences in personality traits, persuasion susceptibility, and emotional intelligence (e.g., StP-II-B, BFI-2-XS, TEIQue-SF) are expected to predict both (a) baseline and post-training performance and (b) subjective reactions to the training. 

Together, these hypotheses enable a systematic assessment of whether, and under which conditions, personalization and duration improve the effectiveness of LLM-generated phishing training.

\subsection{Material, procedure, and measures}
We reused the same experimental infrastructure as in Study~1. The email dataset, web platform, psychometric questionnaires, and five-phase procedure (pre-test, training, post-test, and post-study questionnaire) were identical. 

The only differences concerned the training modules themselves, which were now instantiated in four variants crossing \textit{personalization} (generic vs.\ personalized) and \textit{training length} (short vs.\ long). In all conditions, the content was generated by the LLM using the same modular structure (Introduction, Scenario, Defense Strategies, Exercises, Conclusion), while varying whether user profiles were incorporated (personalized vs.\ generic) and how much detail and number of examples were provided (short vs.\ long).

\subsection{Results}

This section presents the results of Study~2, organized around the four hypotheses concerning the effects of \emph{personalization} and \emph{training length} in LLM-generated phishing-awareness training. First, we summarize descriptive patterns and the overall training effect, and then address each hypothesis in turn. These patterns are visually summarized in Figure~\ref{fig:study1_prepost_box}.

To ensure that the personalization manipulation was effective, we conducted a qualitative inspection on a random 25\% subset of the generated modules, comparing personalized content against generic baselines. The analysis confirmed that the LLM successfully incorporated psychometric profiles into the content generation process, adapting lexical and structural properties while maintaining consistent core instructional messages. For instance, modules generated for participants with high Needs for Certainty featured authoritative tones and structured checklists, whereas content targeting low Premeditation (high impulsivity) emphasized "stop-and-think" mechanisms with shorter sentences compared to the standard descriptive language used in generic modules. This confirms that the lack of performance differentiation was not due to a failure in content generation, but rather suggests that these stylistic adaptations may not be sufficient to drive immediate behavioral changes in a classification task.

\begin{figure*}[ht]
  \centering
  \includegraphics[width=.9\linewidth]{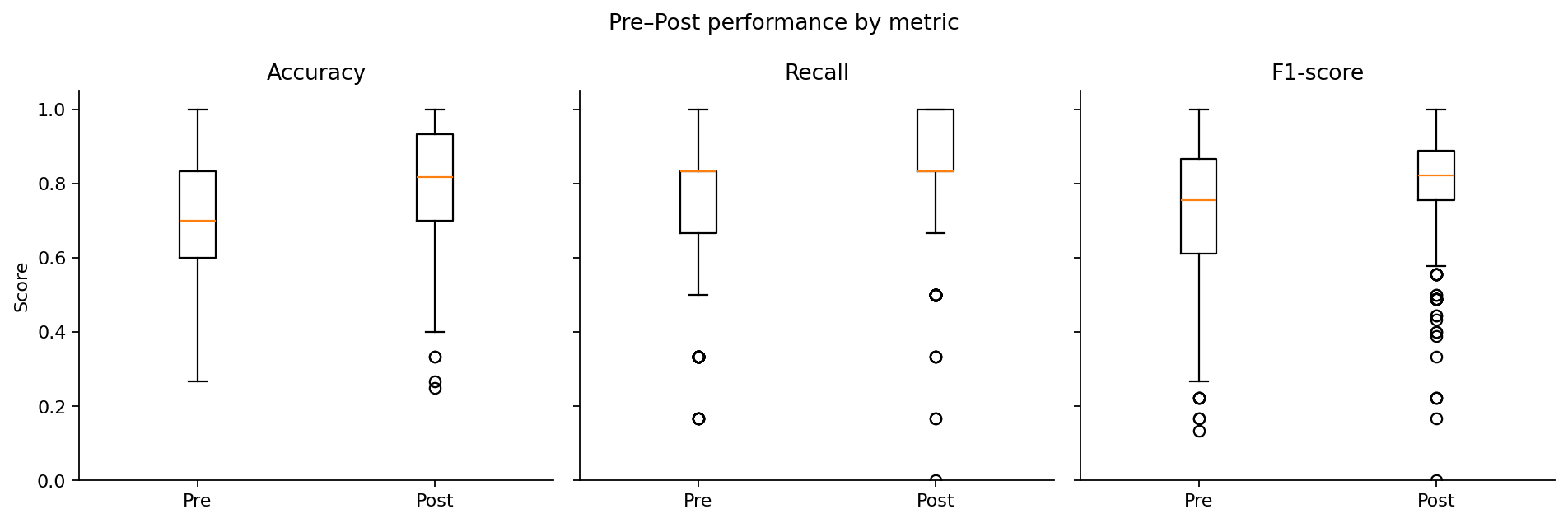}
  \caption{Pre--Post distributions of performance in Study~2. 
  Boxplots show Accuracy, Recall, and F1-score aggregated across the four training conditions, 
  highlighting the systematic shift towards higher scores after training.}
  \label{fig:study2_boxplot_prepost_panel}
\end{figure*}

\subsubsection{Statistical Analysis}
Data analysis was conducted using a combination of variance and correlation techniques, tailored to the specific experimental hypotheses.

To evaluate the impact of training design choices (H1--H5), we employed mixed-design Analysis of Variance (ANOVA). Specifically, we treated \textit{Phase} (Pre vs. Post) as a within-subjects factor and the training configuration (e.g., Length, Personalization, or the four-level Condition factor) as a between-subjects factor. This approach allowed us to assess both the overall learning effect (main effect of Phase) and whether specific training variants produced superior gains (interaction effects). Significant interactions or main effects were followed up with pairwise comparisons. To ensure robust inference for these confirmatory analyses, we applied the \textit{Holm--Bonferroni} correction to control the Family-Wise Error Rate (FWER).

To investigate the role of individual differences (H6), we computed Pearson product-moment correlations between psychometric trait scores (BFI-2-XS, StP-II-B, TEIQue-SF) and performance metrics or training reactions. Given the exploratory nature of this research question and the high dimensionality of the data (30 predictors $\times$ 9 outcomes), we applied the \textit{Benjamini--Hochberg False Discovery Rate (FDR)} correction. This method offers greater statistical power to detect potential patterns in large-scale exploratory screenings while maintaining a controlled proportion of false discoveries.

\subsubsection{H1: Does AI-generated training improve phishing resilience?}

All the four training conditions (\textit{generic-short}, \textit{generic-long}, \textit{personalized-short}, \textit{personalized-long}) showed clear Pre$-$Post improvements in accuracy, recall, and F1-score. As summarized in Table~\ref{tab:desc_study2}, Post-training scores were consistently higher than Pre-training scores in every condition. For instance, the average accuracy increased from approximately $0.70$ to $0.73$ at the Pre-Test to around $0.78$ to $0.84$ at the Post-Test, with similar upward shifts for recall and F1-score.

A mixed ANOVA with Phase (Pre vs.\ Post) as the within-subjects factor and Training Condition (4-level) as the between-subjects factor confirmed a robust main effect of Phase across all metrics. Participants showed statistically significant improvements in \textbf{Accuracy} ($F(1, 394) = 87.81, p < .001, \eta^2 = .182$), \textbf{Recall} ($F(1, 394) = 103.89, p < .001, \eta^2 = .209$), and \textbf{F1-score} ($F(1, 394) = 75.43, p < .001, \eta^2 = .161$). Conversely, neither the main effect of Training Condition nor the Phase $\times$ Condition interaction reached statistical significance for any dependent variable (all $F \le 2.30, p \ge .077$), indicating that while participants learned reliably over time, the four training variants did not differ systematically in overall performance trajectories when treated as distinct prompting groups. Detailed ANOVA tables are provided in the supplementary material.

Conceptually, these findings replicate the \textit{training effectiveness} pattern documented in Study 1 (H1): LLM-generated training produced robust pre-post performance gains irrespective of the specific prompting strategy. On this basis, we focused the hypothesis tests on the factors distinguishing the variants: personalization and training length. Regarding duration, a mixed ANOVA revealed a small but statistically significant main effect of \textbf{Length} on \textbf{Accuracy} ($F(1, 396) = 3.92, p = 0.048, \eta^2 = 0.010$), whereas no significant effects were observed for \textbf{Recall} ($F(1, 396) = 0.50, p = 0.481$) or \textbf{F1-score} ($F(1, 396) = 1.73, p = 0.189$). In contrast, personalization did not yield measurable performance differences across any metric.

\begin{table*}[b]
\centering
\caption{Descriptive statistics (Study~2) by training condition and phase. 
Values are means with standard deviations in parentheses.}
\label{tab:desc_study2}
\begin{tabular}{llccc}
\toprule
Condition & Phase & Accuracy $M$ (SD) & Recall $M$ (SD) & F1 $M$ (SD) \\
\midrule
Generic--short  & Pre  & 0.696 (0.154) & 0.740 (0.211) & 0.704 (0.182) \\
               & Post & 0.798 (0.153) & 0.869 (0.179) & 0.802 (0.165) \\
\midrule
Personalized--short & Pre  & 0.709 (0.165) & 0.742 (0.211) & 0.716 (0.181) \\
                   & Post & 0.775 (0.149) & 0.839 (0.180) & 0.788 (0.147) \\
\midrule
Generic--long   & Pre  & 0.732 (0.158) & 0.743 (0.215) & 0.727 (0.183) \\
               & Post & 0.838 (0.117) & 0.888 (0.140) & 0.842 (0.118) \\
\midrule
Personalized--long & Pre  & 0.728 (0.179) & 0.752 (0.207) & 0.727 (0.196) \\
                  & Post & 0.781 (0.152) & 0.850 (0.186) & 0.786 (0.163) \\
\bottomrule
\end{tabular}
\end{table*}

\subsubsection{H2: Does Personalization Improve Training Effectiveness?}

Hypothesis~2 proposed that \emph{personalized} training would lead to higher Post-training performance than \emph{generic} training. In Study~2, personalization was operationalized by contrasting the two \textit{personalized} conditions with the two \textit{generic} conditions (generic).

Descriptively, personalized and generic variants achieved very similar improvements. Change scores (Post$-$Pre) for accuracy, recall, and F1-score (details in the supplementary material) showed positive gains in all four conditions, but the magnitude of these gains did not consistently favor personalized training. In fact, the generic conditions sometimes exhibited slightly larger mean deltas than the personalized ones, particularly in terms of accuracy and F1-score, although these differences were small in absolute terms.

\begin{figure*}[t]
  \centering
  \includegraphics[width=.95\linewidth]{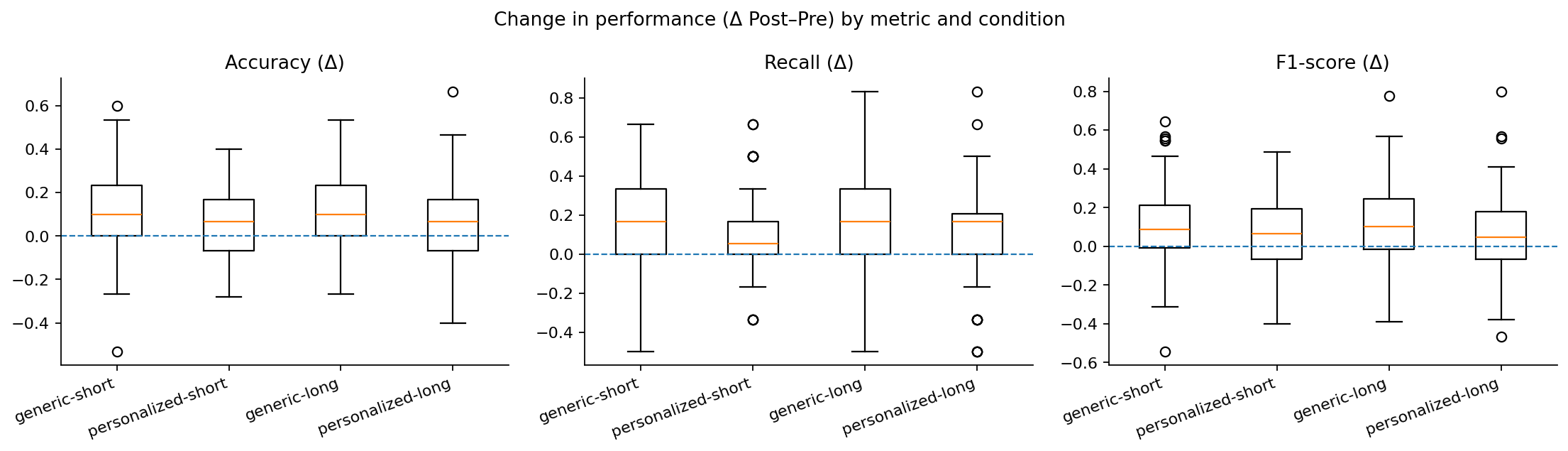}
  \caption{Change in performance ($\Delta = \text{Post} - \text{Pre}$) by training condition 
  and metric in Study~2. Boxplots show improvements in Accuracy, Recall, and F1-score for 
  the four training variants (generic short/long, personalized short/long).}
  \label{fig:study2_boxplot_delta_panel}
\end{figure*}

Figure~\ref{fig:study2_boxplot_delta_panel} summarizes the distribution of improvement scores 
across the four training conditions for all three metrics, illustrating that gains are broadly 
comparable between personalized and generic variants.

Formal analyses corroborated this impression. A one-way ANOVA on improvement scores ($\Delta = \text{Post} - \text{Pre}$) across the four training conditions revealed no statistically significant differences for \textbf{Accuracy} ($F(3, 394) = 2.25, p = 0.082, \eta^2 = 0.017$), \textbf{Recall} ($F(3, 394) = 1.04, p = 0.376, \eta^2 = 0.008$), or \textbf{F1-score} ($F(3, 394) = 1.68, p = 0.171, \eta^2 = 0.013$). Consistent with this result, pairwise post-hoc comparisons between generic and personalized variants failed to reach significance after Holm correction (all $p_{\text{Holm}} \ge .221$).

Taken together, these findings indicate that, in this study, \emph{personalization did not produce measurable benefits} over generic training. If anything, the descriptive pattern suggests that the current implementation of profile-based personalization (via the direct-profile strategy) is at best neutral and does not yield reliable performance gains beyond those obtained with well-designed generic content.

\subsubsection{H3: Does Training Length Improve Performance?}

Hypothesis~3 predicted that \emph{long} training would outperform \emph{short} training, under the assumption that extended interventions provide more examples, explanations, and opportunities for reflection on phishing cues.

To isolate this effect, we collapsed across personalization and compared short vs.\ long versions of the training using a mixed ANOVA with \textit{Length} (Short vs.\ Long) as the between-subjects factor and \textit{Phase} (Pre vs.\ Post) as the within-subject factor. As in the previous analysis, \textbf{Phase} showed a strong main effect for all metrics, confirming robust learning gains over time: \textbf{Accuracy} ($F(1, 396) = 86.78, p < .001, \eta^2 = 0.180$), \textbf{Recall} ($F(1, 396) = 103.64, p < .001, \eta^2 = 0.207$), and \textbf{F1-score} ($F(1, 396) = 74.85, p < .001, \eta^2 = 0.159$).

While the overall learning effect was substantial, the effect of length was more nuanced. For \textbf{accuracy}, there was a statistically significant but very small main effect of length, $F(1,396)=3.92$, $p=.048$, $\eta^2=.010$, indicating that, when aggregating across personalization, long-training participants tended to reach slightly higher accuracy than short-training participants. For \textbf{recall} and \textbf{F1-score}, however, the main effect of length did not reach significance ($F(1,396)=0.50$, $p=.481$, $\eta^2=.001$ for recall; $F(1,396)=1.73$, $p=.189$, $\eta^2=.004$ for F1), and no length~$\times$~phase interaction emerged for any metric.

Descriptive patterns align with these results. Across personalization conditions, Post-training scores were consistently higher for long than for short training, though numerical advantages were modest ($\approx 0.02$--$0.03$ points). Within-condition paired comparisons confirmed that all four experimental groups benefited significantly from the intervention, with large effect sizes. Specifically, the \textbf{Generic-Long} condition exhibited the strongest standardized gains (Hedges' $g = 0.76$ for Accuracy, $0.80$ for Recall), whereas the \textbf{Personalized-Long} condition showed comparatively smaller but still robust improvements ($g = 0.32$ for Accuracy, $0.50$ for Recall). All pre-post differences were statistically significant ($t > 2.82, p \le .006$), confirming the baseline effectiveness of the generated content across all variations.

Overall, these findings provide \emph{partial support} for H3. Longer training reliably improves performance, but its advantage over short training is statistically robust only for accuracy and is modest in magnitude. For recall and F1-score, the evidence for a length benefit is descriptive rather than inferentially strong.

\subsubsection{H4: Do Personalization and Length Interact?}

Hypothesis~4 asserted a positive interaction between personalization and length, such that personalization benefits would be amplified in the long condition, where more detailed and individualized content could be delivered.

The observed pattern does not support this prediction. Examination of improvement scores ($\Delta$) reveals that the \textbf{Generic-Long} condition consistently yielded the largest gains across all metrics ($\Delta\text{Acc} = 0.107, \Delta\text{Rec} = 0.145, \Delta\text{F1} = 0.115$). Contrary to the interaction hypothesis, the \textbf{Personalized-Long} condition did not outperform the others; in fact, it exhibited the smallest improvements in Accuracy ($\Delta = 0.053$) and F1-score ($\Delta = 0.058$), lagging behind even the short generic variant. This trend confirms the absence of a synergistic effect between training duration and the current personalization strategy.

Because the factorial ANOVA with an explicit \textit{Personalization} factor was not retained, interaction tests relied on direct comparisons of the improvement scores ($\Delta$) across the four conditions. As noted, the omnibus ANOVA revealed no significant main effect ($F \le 2.25, p \ge .082$). Crucially, pairwise post-hoc contrasts confirmed that the \textbf{Personalized-Long} condition did not differ significantly from any other group after Holm correction (all $p_{\text{Holm}} \ge .221$). Specifically, it failed to outperform its direct counterpart, \textbf{Generic-Long} ($t = 2.10, p_{\text{Holm}} = .221$ for Accuracy), indicating that the pattern of gains does not support the hypothesized synergy between personalization and training duration.

Taken together, these results suggest that \emph{the benefits of longer training do not depend on whether the content is personalized in the way implemented here}. Length and personalization appear to operate largely independently, with length exerting a modest positive effect (mainly on accuracy) and personalization contributing little additional variance.

\subsubsection{H5: Do Subjective Reactions Mirror Objective Performance?}

Hypothesis~5 predicted that personalization and longer training would lead to more positive subjective evaluations, such as higher perceived usefulness, engagement, and satisfaction. To test this, we analyzed six questionnaire items capturing participants' reactions to the training.

Descriptively, ratings were high across the board. Mean scores for all items and conditions ranged approximately between $3.2$ and $4.3$ on a 1--5 scale, indicating that participants generally perceived the training as useful, clear, and engaging, regardless of the specific variant they received. There was a mild tendency for long, generic training to obtain slightly higher means on some items (e.g., perceived thoroughness or clarity), but these differences were small.

\begin{figure*}[t]
  \centering
  \includegraphics[width=.9\linewidth]{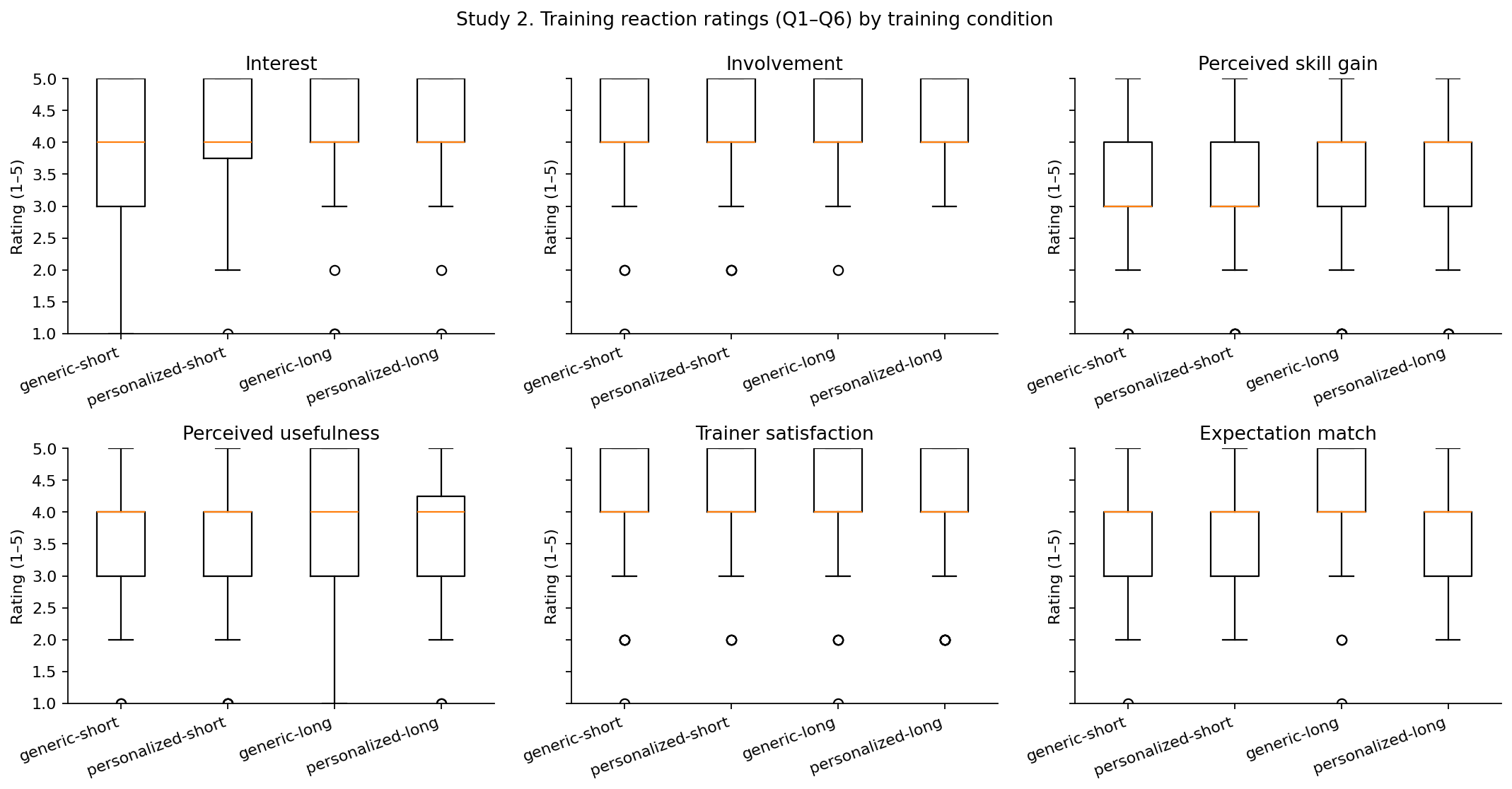}
  \caption{Study~2. Boxplots of training reaction ratings (Q1--Q6) across the four training conditions (Generic short, Generic long, Personalized short, Personalized long). Each panel corresponds to one questionnaire item, with higher values indicating more positive evaluations.}
  \label{fig:study2_qitems_boxplots}
\end{figure*}

Figure~\ref{fig:study2_qitems_boxplots} provides a visual summary of participants' reactions to the training across the four conditions. Across all six items (Q1--Q6), the distributions are tightly clustered, with median ratings generally ranging from 3.5 to 4.5 on the 1--5 scadespite performance data suggestings overlap: one-way ANOVAs performed on each item with \textit{Prompting} (4-level) as the factor yielded no statistically significant main effects (all $F(3, 394) \le 2.31, p \ge .076, \eta^2 \le .017$). This lack of differentiation suggests that participants' subjective evaluations---including perceived usefulness and satisfaction---were uniformly positive and independent of the specific training variant received.

Thus, \emph{subjective reactions did not mirror the modest objective advantages of longer training} and showed no evidence of a personalization benefit. Participants appeared to evaluate all four training variants similarly, despite performance data suggesting slightly higher accuracy in the long conditions. This dissociation is consistent with prior work in security training, where users often struggle to gauge which interventions are most effective accurately.

\subsubsection{H6: Do psychometric factors correlate with performance and training?}

To explore whether individual differences predicted performance or the subjective experience of the training, we correlated all StP-II-B, BFI-2-XS, and TEIQue-SF scales with (a) pre-, post-, and delta performance (accuracy, recall, F1) and (b) participants’ reactions to the training (Q1--Q6). 
All analyses used Pearson correlations with Benjamini--Hochberg FDR correction.

\paragraph{Performance (Pre, Post, and $\Delta$).} Across the psychometric predictors and 9 performance outcomes, \emph{no psychometric scale robustly predicted improvement scores} ($\Delta$ accuracy, recall, F1); all adjusted $p$-values exceeded $.05$. This suggests that the amount of learning generated by the LLM-generated training was largely independent of participants' cognitive and affective dispositions.

Several small but consistent relationships emerged for baseline and post-training performance levels. In particular, \textit{positive attitudes toward advertising} (StP-II-B) showed reliable \emph{negative} correlations with accuracy, recall, and F1 at both Pre and Post (e.g., $r$'s between $-.18$ and $-.21$, all $p_{\mathrm{FDR}} < .01$). 
Related persuasion-related traits---such as \textit{need for uniqueness}, \textit{social influence}, and \textit{lack of self-control}---also showed weak negative associations with performance across multiple metrics ($|r| \approx .10$--$.16$, surviving FDR correction in some cases).
Conversely, no BFI-2-XS or TEIQue-SF scale demonstrated robust associations with performance after correction, suggesting that personality and \ac{EI} had at most a minor influence on task accuracy.

\begin{table*}[t]
\centering
\caption{Significant psychometric correlates of performance (Study~2). 
All correlations are Pearson $r$ values surviving FDR correction ($p_{\mathrm{FDR}} < .05$).}
\label{tab:corr_perf_study2}
\small
\begin{tabular}{@{}llcc@{}}
\toprule
Outcome & Predictor & $r$ & $p_{\mathrm{FDR}}$ \\
\midrule
Accuracy (Pre)  & StP-II-B: Positive attitudes toward advertising & $-0.19$ & $<.010$ \\
Accuracy (Post) & StP-II-B: Positive attitudes toward advertising & $-0.21$ & $.004$ \\
F1 (Post)       & StP-II-B: Positive attitudes toward advertising & $-0.19$ & $.010$ \\
F1 (Post)       & StP-II-B: Need for uniqueness                 & $-0.16$ & $.035$ \\
F1 (Post)       & BFI-2-XS: Agreeableness                       & $-0.16$ & $.048$ \\
Recall (Post)   & StP-II-B: Need for uniqueness                 & $-0.16$ & $.048$ \\
Recall (Pre)    & StP-II-B: Positive attitudes toward advertising & $-0.18$ & $.010$ \\
\bottomrule
\end{tabular}
\end{table*}

\paragraph{Subjective reactions to training (Q1--Q6).} A different pattern emerged for participants’ reactions to the training. 
Several psychometric traits showed small-to-moderate \emph{positive} associations with ratings of usefulness, engagement, clarity, and satisfaction. 
\textit{Conscientiousness} (BFI-2-XS) was the most consistent predictor, correlating positively with all six reaction items ($r \approx .16$--$.20$, all $p_{\mathrm{FDR}} < .01$). 
Similarly, \textit{positive attitudes toward advertising} (StP-II-B) displayed broad positive associations with training reactions ($r \approx .15$--$.27$, all $p_{\mathrm{FDR}} < .01$), suggesting that participants who are more receptive to persuasive content tended to evaluate the training more favorably. 
Additional correlates included \textit{need for uniqueness}, \textit{need for consistency}, and TEIQue-SF components such as \textit{well-being} and \textit{emotionality}. 

Notably, none of these traits predicted learning gains, indicating a clear dissociation between \emph{objective effectiveness} and \emph{subjective appreciation}: participants' impressions of the training were shaped by their dispositions, whereas actual improvement was not.

\begin{table*}[b]
\centering
\caption{Significant psychometric correlates of training reactions (Study~2). 
Only effects with $p_{\mathrm{FDR}} < .05$ are shown.}
\label{tab:corr_react_study2}
\begin{tabular}{@{}llcc@{}}
\toprule
Item & Predictor & $r$ & $p_{\mathrm{FDR}}$ \\
\midrule
Q1--Q6 & BFI-2-XS: Conscientiousness & $0.16$--$0.20$ & $<.01$ \\
Q1--Q6 & StP-II-B: Positive attitudes toward advertising & $0.15$--$0.27$ & $<.01$ \\
Q2--Q5 & StP-II-B: Need for uniqueness & $0.17$--$0.18$ & $<.01$ \\
Q3--Q4 & StP-II-B: Need for avoidance of similarity & $0.16$--$0.18$ & $<.01$ \\
Q4--Q6 & TEIQue-SF: Well-being / Emotionality & $0.13$--$0.15$ & $<.03$ \\
Q3     & StP-II-B: Social influence         & $0.20$ & $.002$ \\
\bottomrule
\end{tabular}
\end{table*}

\paragraph{Exploratory correlations by training condition.} 
To investigate whether psychometric factors predicted performance or training reactions depending on the type of training received, we conducted the same correlation analyses separately for each of the four Study~2 conditions (generic--short, personalized--short, generic--long, personalized--long), again applying FDR correction within each condition.

Across the four conditions, the pattern was remarkably consistent: psychometric profiles did not reliably predict pre--post learning gains ($\Delta$ accuracy, recall, F1) in any arm.
Only one correlation survived FDR correction: in the \textit{personalized-long} condition, \textit{social influence} (StP-II-B) was negatively associated with accuracy improvement ($r=-.39$, $p_{\mathrm{FDR}}=.013$), suggesting that participants reporting higher susceptibility to social pressure tended to improve slightly less. No analogous effects emerged in the other conditions or for any other performance metric; therefore, this isolated finding should be interpreted cautiously.

A more structured pattern emerged for \emph{subjective reactions}, but only in the two personalized conditions. In the \textit{personalized--short} condition, participants with a stronger \textit{need for uniqueness} or more \textit{positive attitudes toward advertising} provided more favorable evaluations of the training, whereas those with a higher \textit{need for avoidance of similarity} rated it less positively on specific items. In the \textit{personalized-long} condition, \textit{need for uniqueness} and \textit{positive attitudes toward advertising} again exhibited positive associations with selected reaction items.

Crucially, none of these profile--reaction associations were present in the generic conditions, and none predicted performance improvements. These results suggest that although psychometric dispositions do not significantly influence training effectiveness, they can modestly shape how participants experience personalized LLM-generated training, consistent with theories of responsiveness to tailored persuasive content ~\cite{kaptein2015personalizing, busch2013personalized}.

A concise summary of all condition-specific significant correlations is provided in Table~\ref{tab:psych_by_condition}. 
A comprehensive table, including all tested pairs, is reported in the supplementary material.

\begin{table*}[b]
\centering
\caption{Significant psychometric correlates by training condition (Study~2).
Only correlations surviving FDR correction within each condition are shown.}
\label{tab:psych_by_condition}
\begin{tabular}{@{}llllcc@{}}
\toprule
Domain & Condition & Outcome & Predictor & $r$ & $p_{\text{FDR}}$ \\
\midrule
Performance & Personalized--Long 
            & $\Delta$ Accuracy 
            & StP-II-B: Social influence 
            & $-0.39$ & .013 \\
\midrule
Reactions   & Personalized--Short 
            & Q3 
            & StP-II-B: Need for avoidance of similarity 
            & $-0.34$ & .030 \\
Reactions   & Personalized--Short 
            & Q3 
            & StP-II-B: Need for uniqueness 
            & $0.34$ & .030 \\
Reactions   & Personalized--Short 
            & Q5 
            & StP-II-B: Positive attitudes toward advertising 
            & $0.33$ & .030 \\
\midrule
Reactions   & Personalized--Long 
            & Q2 
            & StP-II-B: Need for uniqueness 
            & $0.34$ & .037 \\
Reactions   & Personalized--Long 
            & Q3 
            & StP-II-B: Positive attitudes toward advertising 
            & $0.41$ & .003 \\
\bottomrule
\end{tabular}
\end{table*}

\subsubsection{Summary}

Study 2 provides a comprehensive assessment of how personalization and training length impact the effectiveness of LLM-generated phishing-awareness training. Across all analyses, the central findings were highly consistent:

\begin{itemize}
    \item \textbf{H1 (Training effectiveness for phishing resilience)}: \emph{Supported}. This is the most important hypothesis of the entire work, as it confirms the benefit of using \acp{LLM} at scale for generating PETA programs that can improve user learning in phishing attacks. Indeed, post-training performance consistently outperformed that of pre-training, demonstrating the potential of employing \acp{LLM} to enhance users' ability to distinguish between phishing and genuine emails. 
         
    \item \textbf{H2 (Personalization effect)}: \emph{Not supported}. 
    Personalized (profile-based) training did not improve post-training performance relative to 
    generic variants. Learning gains were statistically indistinguishable across personalized and 
    non-personalized conditions, and no psychometric trait reliably moderated the effectiveness of 
    personalization.  
    
    \item \textbf{H3 (Length effect)}: \emph{Partially supported}. 
    Longer training produced a small but statistically reliable improvement in \emph{accuracy}, 
    though benefits for \emph{recall} and \emph{F1-score} were descriptive and did not reach 
    significance. These results were stable after re-analysis with the corrected code and remained 
    consistent across psychometric profiles, none of which predicted larger learning gains.  
    
    \item \textbf{H4 (Interaction)}: \emph{Not supported}. 
    The hypothesized synergy between personalization and length did not emerge.  
    The personalized--long condition did not outperform the generic--long variant, and no interaction 
    pattern was detectable in either raw scores or change scores.  
    Condition-specific correlation analyses confirmed this independence: only a single weak 
    association emerged (social influence negatively predicting $\Delta$accuracy in the personalized--long 
    arm), and it did not generalize to other conditions or metrics.
    
    \item \textbf{H5 (Subjective experience)}: \emph{Not supported}. 
    Participants evaluated all training variants positively, with no systematic differences between
    personalized vs.\ generic or long vs.\ short conditions.  
    Psychometric analyses revealed that subjective reactions were shaped instead by stable individual 
    dispositions---especially \emph{conscientiousness}, \emph{positive attitudes toward advertising}, 
    and \emph{need for uniqueness}---but these traits did not predict actual learning gains.
    
    \item \textbf{H6 (Psychometric factors)}: \emph{Partially supported}. 
    Across all conditions, psychometric traits showed only \emph{weak} associations with baseline and 
    post-training performance (e.g., higher \emph{positive attitudes toward advertising}, 
    \emph{need for uniqueness}, or \emph{agreeableness} were linked to slightly poorer accuracy, recall, 
    and F1), and \emph{no} scale robustly predicted improvement scores ($\Delta$ accuracy, recall, F1). 
    In contrast, several traits---notably \emph{conscientiousness}, \emph{positive attitudes toward 
    advertising}, \emph{need for uniqueness}, \emph{need for avoidance of similarity}, and TEIQue-SF 
    dimensions such as \emph{well-being} and \emph{emotionality}---were consistently associated with 
    more positive subjective reactions, particularly in the personalized arms. 
    A few condition-specific effects emerged (e.g., social influence negatively predicting 
    $\Delta$accuracy in the personalized--long condition), but these were isolated and did not alter 
    the overall pattern that psychometric profiles shape \emph{how the training is experienced} rather 
    than \emph{how much participants learn}.
\end{itemize}

Overall, Study~2 indicates that, in the present implementation, 
\emph{training length matters modestly for objective performance}, whereas 
\emph{static profile-based personalization offers no measurable behavioral advantage}.  
Psychometric profiles influence how participants \emph{feel} about the training, but not how much they \emph{learn} from it. 

These findings suggest that effective personalization for phishing-awareness training may require more adaptive, dynamic, or interaction-based approaches rather than simply injecting the static psychometric profiles into prompting workflows.

\begin{table}[t]
\centering
\caption{Mixed ANOVA (Study~2) with \textit{Phase} (Pre vs.\ Post) as within-subject factor 
and \textit{Length} (Short vs.\ Long) as between-subject factor.}
\label{tab:anova_length_study2}
\small
\begin{tabular}{llcccc}
\toprule
DV & Effect & $F$ & $df_1,df_2$ & $p$ & $\eta^2$ \\
\midrule
Accuracy & Length & 3.92 & 1,396 & .048 & .01 \\
Accuracy & Phase  & 86.78 & 1,396 & $<.001$ & .18 \\
\midrule
Recall   & Length & 0.50 & 1,396 & .481 & .001 \\
Recall   & Phase  & 103.64 & 1,396 & $<.001$ & .21 \\
\midrule
F1       & Length & 1.73 & 1,396 & .189 & .004 \\
F1       & Phase  & 74.85 & 1,396 & $<.001$ & .16 \\
\bottomrule
\end{tabular}
\end{table}

\section{Discussion}

Across two complementary studies, we investigated whether and how LLM-generated phishing-awareness training can effectively improve users’ detection performance, and which design parameters are most critical. Study 1 (four personalized prompting strategies, $N=80$) served as a preliminary screening stage, whereas Study~2 (four prompting/length configurations, $N=400$) provided a higher-powered test of training effects and design trade-offs. In this section, we first summarise the main empirical findings and then discuss their implications for the design of LLM-based phishing-awareness interventions.

\subsection{Benefits of AI-generated training for phishing resilience}

This research provides empirical validation for the feasibility and effectiveness of using AI to generate scalable phishing training. Across two studies involving 480 participants, our results provide converging evidence that generative AI functions as a robust ``instructional engine'': both Study 1 and Study 2 demonstrated significant and substantial pre-post learning gains, regardless of the specific prompting strategy employed. The consistency of these gains across disparate conditions—ranging from simple to complex prompting strategies and varying durations—indicates that modern LLMs possess an inherent capability to structure effective pedagogical narratives without requiring hyper-specialized prompt engineering.

These findings strongly support \textit{H1 (Training effectiveness for phishing resilience)}, confirming that LLMs can produce the core components of effective security training, including realistic scenarios, actionable defense strategies, and immediate feedback. Crucially, the generated content was sufficient to drive significant improvements, particularly in \textit{Recall} (identifying phishing emails), suggesting that the AI-generated training successfully helped users internalize specific threat indicators rather than merely guessing.

This result has profound implications for the scalability of \ac{PETA} programs. Current manual approaches are often labor-intensive and static~\citep{kumaraguru2010school,reinheimer2020investigation}; in contrast, our findings demonstrate that organizations can leverage these models to generate high-quality variations of training programs with minimal human effort. This capability overcomes the bottleneck of manual content creation, potentially enabling \textit{continuous}, on-demand training cycles that can adapt to emerging threats faster than traditional static libraries.

\subsection{Interpreting the role of prompting and personalization}

Study~1 was deliberately generous to the LLM: all four prompting strategies produced \emph{personalised} training tailored to participant profiles. Within this space, the \emph{direct-profile} condition emerged as the most promising configuration in practical terms, achieving the highest Post scores and the largest (or near-largest) Pre--Post gains across all three metrics. However, the lack of significant differences in the ANOVAs and post-hoc tests suggests that this advantage is descriptive rather than statistically significant. A pragmatic reading is that \emph{the model is capable of producing reasonably effective training across a variety of prompting schemes, provided that a minimal level of clarity and structure is maintained}, and that heavy prompt engineering may yield diminishing returns once a baseline of clarity and structure is met.

In Study 2, we moved beyond this preliminary screening and introduced conditions where the LLM generated \emph{non-personalized} training (generic-short/long) alongside personalized variants (personalized-short/long). At the level of the four-condition Prompting factor, there was again no significant main effect, and the ANOVAs on $\Delta$ scores did not reveal reliable differences between conditions. Descriptively, the long non-personalised training actually achieved the highest Post scores and the largest change scores, slightly outperforming the personalised long condition. Taken together, these patterns indicate that \emph{static, profile-based personalisation---as implemented here---does not yield robust incremental benefits beyond those of well-designed, generic LLM-generated training}. 

This does not mean that personalisation is irrelevant in principle, but rather that the specific form we tested (one-shot profile conditioning at generation time) may be too weak or too coarse-grained to produce a clear behavioural advantage over generic content. Another possible explanation is that the psychometric instruments used to generate the profile — while validated and widely adopted — may not capture the dimensions that are most relevant for shaping individualized phishing-awareness content. Different questionnaires, as well as more dynamic or interactive forms of personalization (e.g., adapting examples and explanations in response to users’ ongoing errors or expressed uncertainties), should be investigated to understand the full potential of tailored phishing-awareness interventions.

Consistent with this interpretation, the correlational analyses in Study 2 showed that psychometric traits associated with persuasion sensitivity (e.g., \textit{positive attitudes toward advertising}, \textit{need for uniqueness}) were negatively related to performance but did not predict learning gains. This further suggests that the profile dimensions used for conditioning the LLM may not align with the cognitive mechanisms underlying phishing detection, thereby limiting the potential impact of static one-shot personalization adopted in this study.

\subsection{Length as a design lever: small effects, consistent trends}

The manipulation of training length in Study~2 was motivated by practical constraints: organisations often face tight time budgets for awareness initiatives, and it is therefore crucial to know whether shorter LLM-generated modules can approach the effectiveness of longer ones. The statistical results suggest a nuanced answer.

On the one hand, the main effect of Length was statistically significant only for Accuracy, accounting for roughly 1\% of the variance, with non-significant effects for Recall and F1. On the other hand, the descriptive patterns are remarkably consistent. Across all four prompting conditions, long variants achieved higher post-training performance metrics compared to their short counterparts, and the largest improvements in Accuracy, Recall, and F1 were observed for the generic-long condition. Even when the short versions performed reasonably well, the long ones tended to “stretch” both the ceiling of achievable performance and the size of the gains.

From a design perspective, this pattern supports a pragmatic recommendation: \emph{when time and attention budgets permit, it is advantageous to deploy longer LLM-generated training}, as it yields small but consistent improvements at no additional human authoring cost. Shorter modules remain defensible when deployment constraints are strict; however, they should be viewed as a trade-off rather than an equivalent alternative.

\subsection{Implications for LLM-based security education}

The combined evidence from Study~1 and Study~2 suggests several implications for the design of LLM-based phishing-awareness training:

\begin{enumerate}
    \item \textbf{LLM-generated training is a robust baseline.} Across two independent samples, pre-post improvements were reliable and of non-trivial magnitude, particularly for \textit{Recall}. This supports the viability of modern LLMs as ``instructional generators'' capable of producing effective security training without extensive human curation.

    \item \textbf{Prompt engineering has limited incremental value in this context.} Once prompts are sufficiently clear and oriented towards explanation and examples, different prompting styles (Few-shot, Table-based, Guideline-based, Direct-profile) produce broadly similar learning outcomes. This finding can help organisations avoid over-investing in prompt optimisation for training generation.

    \item \textbf{Static profile-based personalisation is not a guaranteed win.} The transition from a fully personalised space (Study~1) to a mix of personalised and non-personalised conditions (Study~2) did not reveal strong advantages for personalised variants. Designers should therefore be cautious about assuming that simple profile conditioning will automatically enhance training effectiveness.

    \item \textbf{Longer content is beneficial but not transformative.} Longer LLM-generated training yields a small but consistent edge, especially in Accuracy, without radically altering the qualitative pattern of results. This suggests that LLMs are particularly suited to “cheaply” scaling up the quantity and richness of training content, while human designers focus on higher-level structure and integration.
    
    \item \textbf{Individual differences matter more for subjective reactions than for learning gains.} Study~2 showed that personality and persuasion-related traits did not predict how much participants improved, but they did shape how the training was evaluated. Conscientiousness and positive attitudes toward advertising were associated with more favorable reactions, whereas traits such as need for uniqueness were negatively associated with baseline detection performance. This dissociation implies that designers should not assume that individuals who \emph{like} the training are necessarily those who benefit the most, and suggests opportunities for future adaptive systems to tailor motivational framing or explanation style, rather than relying on static psychometric profiles to personalize content.
\end{enumerate}

\section{Limitations}

Our findings should be interpreted in light of several limitations that lay the groundwork for future research.

First, the experimental design did not include a passive control group. While this precludes isolating the pure ``testing effect'' (improvements due solely to repeated exposure to the task), we contend that the observed gains are largely attributable to the training intervention for three reasons. 
(i) Stimulus Independence: We employed two distinct, non-overlapping datasets for Pre- and Post-test, strictly balanced via the NIST Phish Scale to ensure structural equivalence without item repetition. This design eliminates simple memorization effects. 
(ii) Comparative Focus: Since our primary research question concerned the \emph{relative} efficacy of training variants (e.g., Personalized vs.\ Generic), any baseline testing effect is assumed to be constant across randomized groups, thus preserving the validity of the between-condition comparisons. 
(iii) Effect Magnitude: The observed effect sizes (partial $\eta^2 \in [.16, .21]$) substantially exceed the marginal gains typically associated with mere task familiarization in short-term HCI experiments, suggesting a genuine acquisition of discriminative skills.

Second, Study~1 was designed as an exploratory screening intended to identify the most viable prompting strategy for the subsequent experiment, rather than to provide a definitive ranking of all possible prompting variations. Consequently, the sample size ($N=80$) was not powered to detect small effect sizes between the four prompting conditions. The lack of statistical significance between conditions in Study~1 should be interpreted as evidence of functional equivalence for the purpose of selection, rather than proof that no subtle differences exist.

Third, we intentionally focused on comparing different AI-driven generation strategies (e.g., prompting styles, personalization, length) rather than comparing AI-generated content against a human-authored baseline. We excluded a human benchmark because manual training materials introduce uncontrolled variability depending on the specific pedagogical expertise of the author, which makes standardization difficult. Our objective was to evaluate the scalability and intrinsic optimization of automated content generation, which offers advantages in speed and volume that manual authoring cannot match, regardless of comparative qualitative superiority.

Fourth, our implementation of personalization relied on the static injection of psychometric profile data into the LLM prompt. While manipulation checks confirmed that the model altered the output style based on these inputs, this approach treats the LLM as a ``black box'' and does not guarantee that the generated adaptations align perfectly with established pedagogical theory for those specific traits. Furthermore, this form of personalization is one-shot and static; it does not adapt dynamically to the user's performance or misconceptions during the training session, which may explain the limited impact of personalization observed in Study~2.

Finally, the study was conducted in a controlled environment using a web-based classification task. We acknowledge that this setting lacks the ecological validity of a real-world organizational context where users encounter phishing amidst daily work distractions. However, this controlled setting was necessary to isolate the specific effects of content generation variables (prompting, length, personalization) without the noise of organizational confounders. The primary contribution of this work is not to simulate a perfect attack scenario, but to demonstrate the \emph{scalability} of the approach: our results show that LLMs can generate infinite, psychometrically consistent training variations at near-zero marginal cost, providing a baseline of effectiveness that organizations can deploy rapidly.

\section{Conclusion}

This paper presented the first empirical investigation of LLM-generated phishing-awareness training. Across two controlled studies (encompassing 80 and 400 participants), we investigated how prompting strategies, personalization, and training duration impact users’ ability to detect phishing emails.

Study~1 compared four \textit{prompting strategies} for generating training with \acp{LLM}, showing that significant pre-post improvements in accuracy, recall, and F1-score can be obtained regardless of the specific prompt used. Although no statistical differences emerged between conditions, the \emph{direct-profile} configuration achieved the strongest descriptive performance, suggesting that simple prompting pipelines may be sufficient for generating effective training content.

Study~2 extended these findings by evaluating the roles of \emph{personalization} and \emph{training length}. While all conditions yielded substantial learning gains, only training duration had a measurable effect: longer modules led to improvements in accuracy. In contrast, the static profile-based personalization used here did not outperform generic content, indicating that richer or adaptive personalization techniques may be required to produce meaningful benefits.

Overall, our results demonstrate that LLMs can support scalable and effective phishing-awareness training, reducing the manual effort typically required to design instructional materials. At the same time, the findings highlight key design considerations: content richness appears more important than static personalization, and simple prompting strategies can be surprisingly competitive.

Future work should explore adaptive, performance-driven personalization. Additionally, it aims to explore more interactive training formats and the application of LLMs to facilitate rapid experimentation on training variants. As phishing threats continue to evolve, AI-generated content offers a promising foundation for more flexible, data-driven PETA programs.

\section*{Declaration of generative AI}
During the writing of this paper, the author(s) used \textit{Grammarly Pro} to fix grammatical errors and improve text quality. After using this tool/service, the author(s) reviewed and edited the content as needed and take(s) full responsibility for the content of the published article.

\section*{Acknowledgments}
This work has been supported by the Italian Ministry of University and Research (MUR) and by the European Union-NextGenerationEU, under grant PRIN 2022 PNRR "DAMOCLES: Detection And Mitigation Of Cyber attacks that exploit human vuLnerabilitiES" (Grant P2022FXP5B) CUP: H53D23008140001. 

\section*{Data Availability}
The datasets, analysis scripts, and experimental materials generated during the current study are available in the Figshare repository: \url{https://doi.org/10.6084/m9.figshare.30664793}.
    
\bibliographystyle{unsrt}
\bibliography{bibliography,background}

\end{document}